\documentclass[superscriptaddress,
twocolumn,%preprint,
bibnotes,amsmath,amssymb,aps,pra,floatfix]{revtex4-2}

\usepackage[margin=1in]{geometry}
\usepackage[english]{babel}
\usepackage{xcolor}
\usepackage{ucs}
\usepackage[utf8x]{inputenc}
\usepackage{amsmath}
\usepackage{amssymb}
\usepackage{amsthm}
\usepackage{graphicx}
\usepackage{bm}
\usepackage{bbm}
\usepackage{braket}
\usepackage{empheq}
\usepackage[colorlinks=true]{hyperref}
\definecolor{purp}{rgb}{0.5,0,0.5}

\newcommand{\JILA}{JILA, NIST, and Department of Physics, University of Colorado, Boulder, CO 80309, USA}
\newcommand{\OKLA}{Department of Physics, Oklahoma State University, Stillwater, OK 74078, USA}

\begin{document}

\title{Resonant dynamics of strongly interacting SU($n$) fermionic atoms in a synthetic flux ladder}
\date{\today}

\author{Mikhail Mamaev}
\email{mikhail.mamaev@colorado.edu}
\affiliation{\JILA}
\author{Thomas Bilitewski}
\affiliation{\JILA}
\affiliation{\OKLA}
\author{Bhuvanesh Sundar}
\affiliation{\JILA}
\author{Ana Maria Rey}
\affiliation{\JILA}

\begin{abstract}
{
We theoretically study the dynamics of $n$-level spin-orbit coupled alkaline-earth fermionic atoms with SU($n$) symmetric interactions. We consider three dimensional lattices with tunneling along one dimension, and the internal levels treated as a synthetic dimension, realizing an $n$-leg flux ladder. Laser driving is used to couple the internal levels and to induce an effective magnetic flux through the ladder.
We focus on the dense and strongly interacting regime, where in the absence of flux the system behaves as a Mott insulator with suppressed motional dynamics. At integer and fractional ratios of the laser Rabi frequency to the onsite interactions, the system exhibits resonant features in the dynamics. These resonances occur when interactions help overcome kinetic constraints upon the tunneling of atoms, thus enabling motion. Different resonances allow for the development of complex chiral current patterns. The resonances resemble the ones appearing in the longitudinal Hall resistance when the magnetic field is varied. We characterize the dynamics by studying the system's long-time relaxation behavior as a function of flux, number of internal levels $n$, and interaction strength. We observe a series of non-trivial pre-thermal plateaus caused by the emergence of resonant processes at successive orders in perturbation theory. We discuss protocols to observe the predicted phenomena under current experimental conditions.}
\end{abstract}
\maketitle

\section{Introduction}

Understanding the dynamics of interacting quantum particles on a lattice has been a paradigmatic goal of physics research for many decades. The most popular description for fermionic particles hopping on a lattice with onsite interactions is the canonical Fermi-Hubbard model, which is believed to contain the core requisite ingredients for high-$T_c$ superconductivity~\cite{lee2006highTC}. Extensions with more than two internal levels exhibiting a global SU($n$) symmetry in the interaction term have also been used to describe a broad class of models relevant to solid-state~\cite{coleman2015heavy} and high energy physics~\cite{Hofstetter2007}. Unfortunately, many details of the quantum many-body behavior featured by these systems remain elusive.

Arrays of alkaline earth atoms (AEAs) featuring a large number of long-lived nuclear spin levels in both ground and excited electronic state manifolds, together with SU($n$)-symmetric interactions, are emerging as natural quantum simulators of the SU($n$) Hubbard model~\cite{Cazalilla2014, reyTwoOrbitalSUNMagnetism2010}.
Due to the long coherence times in these systems, it is also possible to emulate the presence of a strong external magnetic field via additional laser drives without suffering from heating effects~\cite{spielmanSeminalTopology2019,goldman2016review,Dalibard2011,zhai2015degenerate,Zhang2018}. These versatile quantum simulation capabilities make AEAs an ideal platform for the exploration of new facets of quantum magnetism and topology relevant across multiple disciplines.

Despite such great opportunities, the exploration of many-body physics with AEAs still remains at an early stage.  Alkaline-earth-atom experiments have targeted the non-interacting regime~\cite{fallaniChiralCurrents2015,fallaniSOCSU2Optical2016,Kolkowitz2017,bromley2018dynamics,song2016spin,song2018observation,song2019observation,hanToblerone2019} or the weakly interacting regime~\cite{zhang2014spectroscopic, bromley2018dynamics,song2020,Sonderhouse2020}, for both of which a Fermi-liquid description can be used. There has been some progress in the strongly interacting limit achieved in deep optical lattices, including the observation of few-body SU$(n)$ orbital physics in isolated lattice sites~\cite{goban2018emergence,scazza2014observation,cappellini2014direct}, of equilibrium Mott insulating phases~\cite{taie2010realization,taie20126,hofrichter2016direct,tusi2021flavourselective}, and of antiferromagnetic correlations~\cite{ozawa2018antiferromagnetic,taie2020observation}. However, the rich interacting quantum magnetic behaviors predicted for SU$(n)$ systems have yet to be seen. One major roadblock for experiments thus far is that preparing equilibrium ground state phases has challenging temperature requirements. At the same time, most theory work has focused specifically on ground state properties. These include momentum distributions~\cite{barbarinoFermionicStrip2016}, chiral currents~\cite{vekuaBosonCurrentReversalGS2015, piraudVortexAndMeissnerBoson2015}, topological phases~\cite{meisnerHofstadterFermiHubbardPhases2020}, Laughlin states~\cite{strinati2017laughlinStates} and Hall insulators~\cite{paredesCurrentsSingleParticleOverview2014, selaQuantumHallChiralCurrentsLattice1D2015}, density-dependent magnetism~\cite{santosDensityDependentMagnetism2015}, charge pumping~\cite{huiCurrentsSUN2015}, and connections to baryon physics~\cite{ghoshAttractiveFermiGasesBaryonSUN2017}. Overall, experimentally amenable settings where non-trivial quantum SU$(n)$ magnetic behaviors emerge at currently accessible conditions are scarce.

In this work we explore quantum quench behaviours of SU($n$) interacting fermions that can readily be realized with current generation optical lattice experiments. Our studies focus on fermionic AEAs with $n$ internal levels trapped in a 3D optical lattice and subject to an effective magnetic flux induced by a laser drive. We consider the case where hopping happens only along one spatial direction, so that the overall system including the laser-coupled internal levels (that act as a synthetic dimension) can be visualized as an $ n$-leg flux ladder. We focus on the out-of-equilibrium dynamics in the strongly interacting limit with one atom per site, where in the absence of an external drive the system is a Mott insulator. Generically, strong onsite repulsion is expected to suppress particle motion because of the high energetic cost of forming a doublon. We nonetheless observe the emergence of motion due to multi-body tunneling resonances at fractional or integer values of the ratio between the laser Rabi frequency and the onsite interactions. We first show the emergence of these resonances for SU($2$) atoms, and extend these arguments to SU($n$), finding a significantly larger number of resonances as $n$ increases. At such resonances particle transport is restored in the presence of a non-zero flux, which gives rise to non-trivial chiral currents along the longitudinal lattice tunneling direction. Although in our case the resonances are caused by interaction-enabled tunneling, they resemble the resonances observed at fractional fillings as a function of magnetic field in the fractional quantum Hall effect.

In addition to chiral currents along the real direction, the resonances induce transport along the synthetic dimension which manifests as dressed spin population dynamics. To further characterize the quantum dynamics we also study the long-time behavior of the system at these resonances. We observe the formation of several pre-thermal plateaus, which we can understand in terms of a hierarchy of higher order resonant tunneling processes. Experiments have explored such resonant effects by studying tilted bosonic systems~\cite{daleyTwoSiteBosonic2019,nagerlMultiSiteResonant2014}, tilted Hubbard models exhibiting non-ergodic behavior~\cite{monikaTiltNonErgodic2021}, and staggered tilted Hubbard models used for gauge field simulation~\cite{panGaugeField2020}. The behavior that we find is in striking contrast to more standard thermalization phenomena, revealing the important role of kinetic constraints in our system.

The system we describe can be realized using current state-of-the-art optical lattice experiments. In particular, the key requirements for experimental observation are i) to prepare a spin polarized gas at near-unit filling, ii) to drive either a direct optical transition between ground and excited clock levels (for $n=2$) or Raman transitions between ground nuclear spin levels (for $n\geq 2$), and iii) to use the same laser set-up to measure coherences and populations. All of these requirements are within reach as reported in recent experimental works~\cite{Sonderhouse2020,fallaniChiralCurrents2015,fallaniChiralCurrents2015,fallaniSOCSU2Optical2016,Kolkowitz2017,song2016spin,song2018observation,song2019observation,zhang2014spectroscopic,hanToblerone2019}. Moreover, although we discuss long time thermalization behavior, many of the predicted resonance features happen at timescales set by the lattice tunneling rate (rather than slow effective interaction scales such as superexchange), and should therefore be visible within current optical lattice coherence times. We provide a detailed discussion of possible implementations and discuss various protocols for realizing spin-orbit coupled driving, as well as methods for preparing/measuring the required initial states. We hope that this investigation stimulates experimental work as well as follow-up theoretical research, and facilitates the observation of unique quantum behaviors featured by strongly interacting SU$(n)$ symmetric fermions.

Section~\ref{sec_SU2} will define the model we study for the simplest case of $n=2$ internal states. Quench dynamics of chiral currents in the system will be compared for the non-interacting and strongly interacting limits. A dressed state picture will be used to characterize resonant points and their allowed tunneling processes. Long-time average values of observables will be compared to conventional thermalization predictions. Section~\ref{sec_SUn} will extend the model to general $n$ and show equivalent quantum quench dynamics of currents and dressed state populations. Section~\ref{sec_Implementation} will discuss experimentally realistic protocols for implementation, state preparation, and measurement.

%%%%%
\section{SU(2) system}
\label{sec_SU2}
%%%%%
%%%
\subsection{Fermi-Hubbard model}
%%%
\begin{figure*}
\includegraphics[width=1\textwidth]{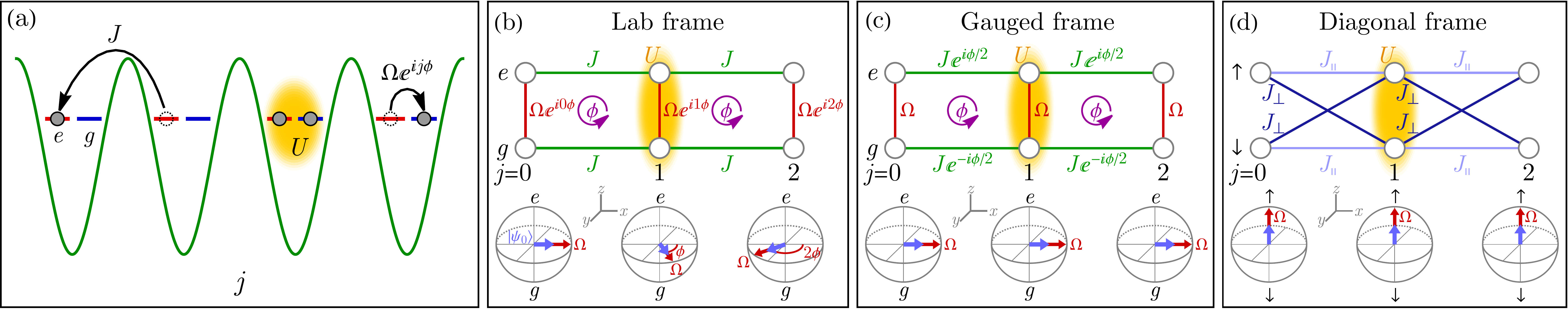}
\centering
\caption{(a) Schematic of the driven optical lattice system for $n=2$ internal spin states labeled by $m \in \{e,g\}$. (b) Effective $L \times 2$ ladder configuration realized by the system in the lab frame, where the spin-orbit coupling phase $e^{i j \phi}$ appears in the laser drive Hamiltonian which couples the internal states. Each plaquette in the 2-leg ladder is pierced by a flux $\phi$. The Bloch spheres at the bottom show the initial product state $\ket{\psi_0}$ considered in this paper with blue arrows, treating each atom as a spin-1/2 degree of freedom assuming a filling fraction of one atom per site. The direction of the drive about which each spin rotates is shown in red. The initial state $\ket{\psi_0}$ is an eigenstate of the drive and points in the same direction. (c) Visualization  of the system in the gauged frame, where the spin-orbit coupling phase is put into the tunneling matrix elements along the lattice direction $j$. (d) Visualization of the system in the diagonal frame where a further basis rotation diagonalizes the drive, defining dressed fermionic states $\nu \in \{\uparrow,\downarrow \}$ as the single-particle eigenstates of the drive with energy $+\Omega/2$, $-\Omega/2$ respectively. The flux enables both spin-conserving and spin-flipping tunneling with rates $J_{\parallel} = J \cos \left(\frac{\phi}{2}\right)$ and $J_{\perp} = J \sin \left(\frac{\phi}{2}\right)$ respectively.}
\label{fig_SU2Schematic}
\end{figure*}

The system we study is a driven three-dimensional optical lattice with tunneling confined to one dimension via strong transverse lattice depths. This arrangement creates an array of independent one dimensional (1D) chains with $L$ sites each. We assume the 1D chains are populated with $N$ fermionic atoms each, prepared in the lowest Bloch band, as depicted in Fig.~\ref{fig_SU2Schematic}(a). The atoms have internal degrees of freedom, which for the moment we restrict to two levels $m \in \{e,g\}$. These can be a pair of electronic clock states or two hyperfine levels.

The Hamiltonian that describes the dynamics of the atoms consists of three main parts,
\begin{equation}
\label{eq_FermiHubbardHamiltonian}
\hat{H} = \hat{H}_{\mathrm{J}} + \hat{H}_{\mathrm{U}} + \hat{H}_{\Omega}.
\end{equation}
The term $\hat{H}_{J}$ describes nearest-neighbour tunneling,
\begin{equation}
\hat{H}_{\mathrm{J}} = -J \sum_{j,m \in \{e,g\}} \left(\hat{c}_{j,m}^{\dagger}\hat{c}_{j+1,m} + h.c.\right),
\end{equation}
where $J > 0$ is the tunneling rate (assuming $\hbar = 1$) and $\hat{c}_{j,m}$ annihilates an atom of spin $m$ on site $j$. The term $\hat{H}_{U}$ is an on-site Hubbard interaction,
\begin{equation}
\hat{H}_{\mathrm{U}} = U \sum_{j} \hat{n}_{j,e}\hat{n}_{j,g},
\end{equation}
where we assume $U > 0$ is a repulsive interaction parameter set by the scattering length of the colliding atoms and $\hat{n}_{j,m} = \hat{c}_{j,m}^{\dagger}\hat{c}_{j,m}$. The last term $\hat{H}_{\Omega}$ is a laser drive that couples the spin states,
\begin{equation}
\hat{H}_{\Omega} = \frac{\Omega}{2} \sum_{j}\left(e^{i j \phi}\hat{c}_{j,e}^{\dagger}\hat{c}_{j,g} + h.c.\right),
\end{equation}
where $\Omega$ is the Rabi frequency (assumed real without loss of generality) and $e^{i j \phi}$ is a Peierls phase imprinted by the laser. Here $\phi=k_L a$ is a differential phase experienced by atoms in adjacent lattice sites, with $a$ the lattice spacing and $k_L$ the magnitude of the drive laser wavevector projected along the lattice direction. This differential phase can be interpreted as a spin-orbit coupling term, since an atom excited from $g \to e$ by absorbing a photon from the laser also acquires an additional momentum $\hbar k_L$. This system is equivalent to a synthetic 2-leg ladder where the sites $j$ along the lattice direction run along the individual legs, and the synthetic spin direction $m$ indexes the legs, see Fig.~\ref{fig_SU2Schematic}(b). The phase $\phi$ amounts to a constant flux piercing each plaquette of this ladder, emulating a transverse magnetic field. 

The Hamiltonian above is written in a ``lab'' frame motivated by experimental implementations. To facilitate better theoretical understanding of the system, we apply a number of basis transformations to simplify the physics. We first make a gauge transformation,
\begin{equation}
\hat{c}_{j,g} = e^{-i j \phi /2}\hat{c}'_{j,g},\>\>\>\hat{c}_{j,e} = e^{+ i j \phi/2}\hat{c}'_{j,e},
\end{equation}
which changes the tunneling to now contain the flux,
\begin{equation}
\begin{aligned}
\hat{H}_{\mathrm{J}}= -J\sum_{j}\big(e^{i \phi/2}&\hat{c}_{j,e}'^{\dagger}\hat{c}'_{j+1,e} \\
+ e^{-i \phi /2}&\hat{c}_{j,g}'^{\dagger}\hat{c}'_{j+1,g} +h.c.\big),
\end{aligned}
\end{equation}
while the drive is now homogeneous across every lattice site,
\begin{equation}
\hat{H}_{\Omega} = \frac{\Omega}{2}\sum_{j}\left(\hat{c}_{j,e}'^{\dagger}\hat{c}'_{j,g}+h.c.\right).
\end{equation}
The interactions remain unchanged. This transformation is analogous to a change of the vector potential (represented here by the spin-orbit coupling phase $e^{i j \phi}$) from a Landau gauge along the synthetic direction to a Landau gauge along the lattice direction. Figure~\ref{fig_SU2Schematic}(c) shows the system in this ``gauged'' frame. A spiral product state in the lab frame with winding angle of $\phi$ per lattice site simplifies to a coherent spin state in the gauged frame and vice versa. Next, we make a second on-site basis rotation,
\begin{equation}
\hat{a}_{j,\uparrow} = \frac{1}{\sqrt{2}}\left(\hat{c}'_{j,g} + \hat{c}'_{j,e}\right),\>\>\>\>\hat{a}_{j,\downarrow} = \frac{1}{\sqrt{2}}\left(\hat{c}'_{j,g} - \hat{c}'_{j,e}\right).
\end{equation}
The new operators $\hat{a}_{j,\uparrow}$, $\hat{a}_{j,\downarrow}$ annihilate atoms in dressed spin states $\nu \in \{\uparrow,\downarrow \}$ that correspond to single-particle eigenstates of the drive with energy $+\Omega/2$, $-\Omega/2$ respectively. Under this transformation the Hubbard interactions still maintain the same form $\hat{H}_{\mathrm{U}} = U\sum_{j}\hat{n}_{j,\uparrow}\hat{n}_{j,\downarrow}$ with $\hat{n}_{j,\nu} = \hat{a}_{j,\nu}^{\dagger}\hat{a}_{j,\nu}$, while the drive becomes diagonal,
\begin{equation}
\hat{H}_{\Omega} = \frac{\Omega}{2}\sum_{j}\left(\hat{n}_{j,\uparrow} - \hat{n}_{j,\downarrow}\right).
\end{equation}
Since a non-zero flux $\phi$ breaks the SU(2) symmetry of $\hat{H}_{\mathrm{J}}$, the tunneling no longer conserves spin and instead both spin-preserving and spin-flipping tunneling terms emerge:
\begin{equation}
\begin{aligned}
\hat{H}_{\mathrm{J}}= -\sum_{j} \bigg[ J_{\parallel} &\left(\hat{a}_{j,\uparrow}^{\dagger} \hat{a}_{j+1,\uparrow} + \hat{a}_{j,\downarrow}^{\dagger}\hat{a}_{j+1,\downarrow}\right)\\
 - i J_{\perp}& \left(\hat{a}_{j,\uparrow}^{\dagger} \hat{a}_{j+1,\downarrow} + \hat{a}_{j,\downarrow}^{\dagger}\hat{a}_{j+1,\uparrow}\right) + h.c.\bigg],
\end{aligned}
\end{equation}
where $J_{\parallel} = J \cos \left(\phi/2\right)$ and $J_{\perp} = J \sin \left(\phi/2\right)$. Fig.~\ref{fig_SU2Schematic}(d) shows the system in this last ``diagonal'' frame. For $\phi = 0$ we only have spin-conserving tunneling $J_{\parallel} = J$, $J_{\perp} = 0$. Maximum $\phi = \pi$ yields only spin-flip tunneling $J_{\perp} = J$, $J_{\parallel}=0$.

Before proceeding further, we clarify that our goal is to study quench dynamics of this system using exact numerical integration of the Schr\"{o}dinger equation. The numerically accessible system sizes $L$ are limited due to the rapid Hilbert space growth. To minimize finite size effects we will use periodic boundary conditions unless otherwise specified. For periodic boundaries, if the flux $\phi$ is an integer multiple of $2\pi/L$, then the laser drive will only induce transitions between well-defined  quasimomentum states. If the flux does not satisfy this requirement (i.e. it is incommensurate), the drive will induce transitions into many different quasimomentum states, complicating the dynamics~\cite{Liang2021}. However, in this work we will be studying the long-time averaged dynamics of collective observables for which an incommensurate $\phi$ will only yield $1/L$ corrections. We thus ignore the issue of commensurability when making numerical comparisons for varying $\phi$. A more thorough discussion of finite size scaling is given in Appendix~\ref{app_Scaling}.

%%%
\subsection{Chirality}
\label{subsec_SU2Chirality}
%%%

The system described by $\hat{H}$ in the lab or gauged frames is a minimal version of a 2D strip pierced by a magnetic field. In the non-interacting case $U = 0$ such a system is described by the Harper-Hofstadter model~\cite{harperHofstadter1955,hofstadter1976HH,spielmanSeminalTopology2019,goldman2016review,Dalibard2011,zhai2015degenerate}, which features chiral currents at its boundaries. While a synthetic dimension of only $n=2$ states is a limiting case of the full two-dimensional Harper-Hofstadter model, one can still gain insights into the dynamical behavior from studying this limit~\cite{paredesCurrentsSingleParticleOverview2014}; we will consider systems with $n>2$ in Section~\ref{sec_SUn}.

\begin{figure}
\includegraphics[width=.49\textwidth]{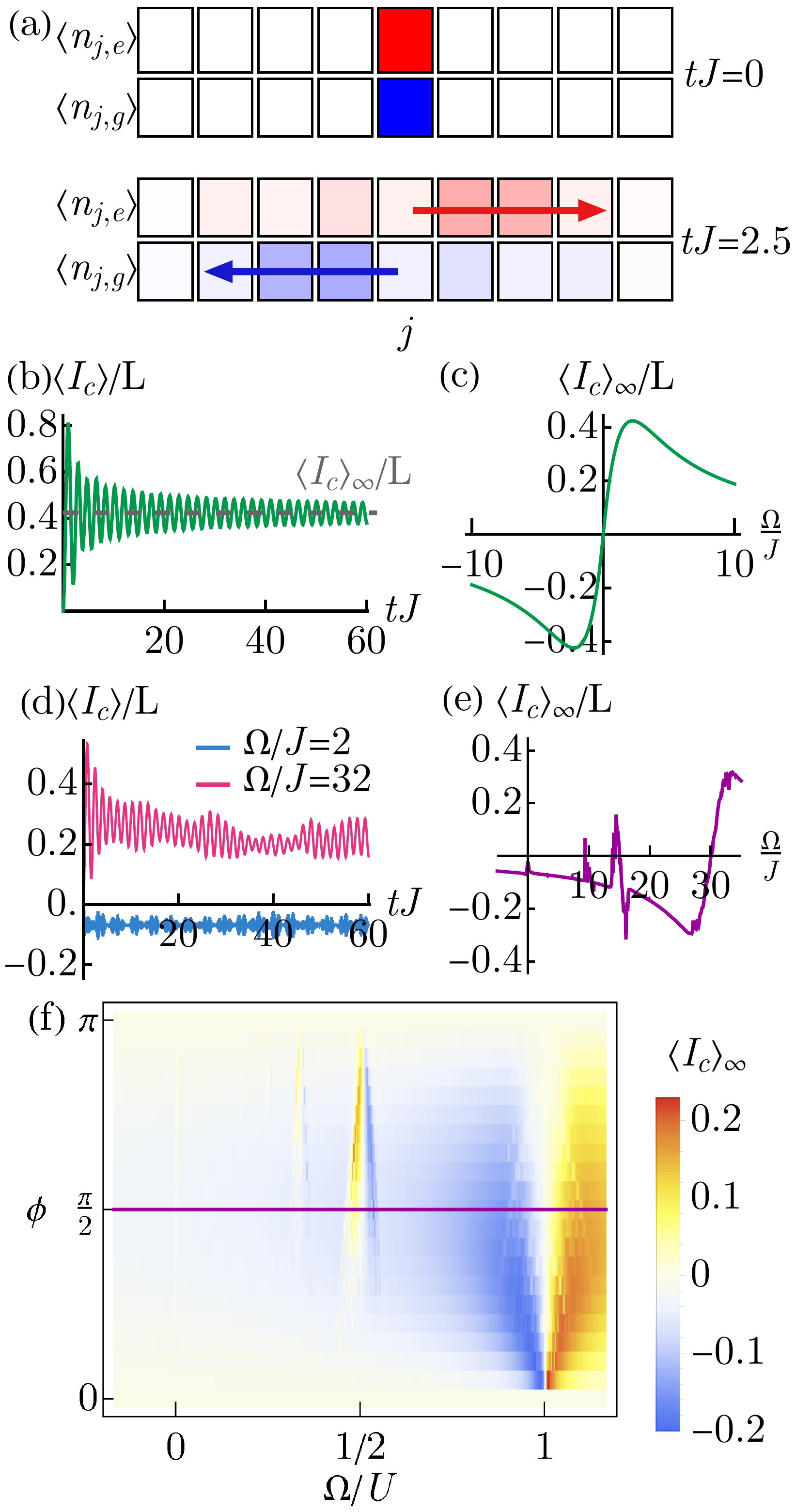}
\centering
\caption{(a) Time-evolution snapshots of populations $\langle \hat{n}_{j,e}\rangle$, $\langle \hat{n}_{j,g} \rangle$ for a single-atom initial state $\ket{\psi_0}_{\text{1 atom}}=\frac{1}{\sqrt{2}}(\hat{c}_{j=0,e}'^{\dagger} + \hat{c}_{j=0,g}'^{\dagger})\ket{0}$ in the gauged frame. The parameters are $\Omega/J = 2$ and $\phi = \pi/2$. (b) Evolution of the chiral current for a non-interacting system $U=0$ with flux $\phi = \pi/2$, drive $\Omega/J=2$, and system size $L\to \infty$. (c) Long-time average current in terms of $\Omega/J$ for the non-interacting system using the same flux. The analytic expression is given in Eq.~\eqref{eq_SU2nonInteractingExact}. (d) Evolution of the chiral current for a strongly interacting system $U/J=30$ with flux $\phi=\pi/2$ and drive strength $\Omega/J=2$, $32$. System size is $L=9$. (e) Average current for the interacting system with varying $\Omega/J$, averaging over $tJ=0$ to $500$. (f) Average current for the strongly interacting system across the parameter space of $\Omega/U$ and $\phi$. The violet line corresponds to panel (e).}
\label{fig_Chirality}
\end{figure}

The magnetic field in the $L \times 2$ strip will induce a shearing effect where $e$-atom population flows in one direction while $g$-atom population flows in the other. A simple example of this shearing is shown in Fig.~\ref{fig_Chirality}(a) for the case of a single atom prepared in an initial superposition state $\ket{\psi_0}_{\mathrm{1 atom}}=\frac{1}{\sqrt{2}}\left(\hat{c}_{j=0,e}'^{\dagger} + \hat{c}_{j=0,g}'^{\dagger}\right)\ket{0}$ with a flux of $\phi = \pi/2$ and a drive strength of $\Omega / J = 2$. The chiral dynamics manifests as the $e$-population $\langle \hat{n}_{j,e} \rangle$ shifting towards lattice sites with increasing $j>0$ while the $g$-population $\langle \hat{n}_{j,g}\rangle$ has an equal and opposite drift towards sites with $j<0$.

The rate of flow of spin population can be quantified via current observables (in units of the tunneling rate $J$),
\begin{equation}
\begin{aligned}
    \hat{I}_{j,e} &=- i \left(\hat{c}_{j,e}^{\dagger}\hat{c}_{j+1,e}-h.c.\right),\\
    \hat{I}_{j,g} &= - i \left(\hat{c}_{j,g}^{\dagger}\hat{c}_{j+1,g}-h.c.\right),
\end{aligned}
\end{equation}
where $\hat{I}_{j,e}$ ($\hat{I}_{j,g}$) corresponds to the rate of transfer of $e$ ($g$) atom population from site $j$ to site $j+1$ per unit time. To quantify the shearing for generic initial conditions we define a bulk chiral current in units of $J$,
\begin{equation}
\begin{aligned}
    \hat{I}_{c} &=\sum_{j} \hat{I}_{j,e} - \sum_{j}\hat{I}_{j,g}\\
    &= -2 \sum_{k} \sin(k a )\left(\hat{n}_{k,e} -\hat{n}_{k,g}\right),
\end{aligned}
\end{equation}
where $\hat{n}_{k,m} = \hat{c}_{k,m}^{\dagger}\hat{c}_{k,m}$ and $\hat{c}_{k,m} = \frac{1}{\sqrt{L}} \sum_{j}e^{i j k a} \hat{c}_{j,m}$. On the second line we have shown that this current can also be obtained from the atoms' quasi-momentum distribution, which is experimentally accessible via e.g. time-of-flight measurements. While such currents have been experimentally observed in optical lattices~\cite{greinerChiralCurrents2017,monikaChiralCurrents2014,fallaniChiralCurrents2015,spielmanChiralCurrents2015,gadwayChiralCurrents2017,Kolkowitz2017}, the role that interactions play in the many-atom limit is still waiting for further exploration, especially in the context of dynamical evolution.

To understand the interplay between the Hubbard interactions and the single-particle physics, we study the quench dynamics of the chiral current for a unit filled system with one particle per site $N/L = 1$. The initial state for the quench we choose is:
\begin{equation}
\ket{\psi_0} = \frac{1}{2^{L/2}} \prod_j \left(\hat{c}_{j,e}'^{\dagger} + \hat{c}_{j,g}'^{\dagger}\right)\ket{0}= \prod_j \hat{a}_{j,\uparrow}^\dagger\ket{0}.
\end{equation}
This state is maximally delocalized along the synthetic ladder. The state can be visualized using a collection of Bloch spheres, each one representing a spin-$1/2$ particle at lattice site $j$. In the lab frame [Fig.~\ref{fig_SU2Schematic}(b)], the initial state is a spiral state with winding angle $\phi$ on the Bloch sphere equator. In the gauged frame [Fig.~\ref{fig_SU2Schematic}(c)] the state instead looks like a uniform collective state along the $+x$ Bloch sphere direction, while in the diagonal frame [Fig.~\ref{fig_SU2Schematic}(d)] it is collective along $+z$. This state is the highest energy eigenstate of the drive $\hat{H}_{\Omega} \ket{\psi_0} = \frac{L \Omega}{2} \ket{\psi_0}$ for $\Omega > 0$, and corresponds to all atoms in the dressed spin state $\uparrow$. Such an initial state can be prepared with an appropriate pulse sequence or adiabatic ramp as discussed in Section~\ref{sec_Implementation}. Note that for $\phi=0$, the system exhibits no dynamics independent of the Hamiltonian parameters $J,U,\Omega$ due to Pauli-blocking.

Fig.~\ref{fig_Chirality}(b) shows a sample time evolution of the chiral current from this initial state, starting with a non-interacting ($U=0$) system for a flux of $\phi = \pi/2$ and drive strength $\Omega/J = 2$. The current undergoes an initial growth and saturates to an average about which it undergoes coherent oscillations. We compute this long-time average $\langle\hat{I}_{c} \rangle_{\infty}$, given by
\begin{equation}
\langle \hat{\mathcal{O}}\rangle_{\infty} = \lim_{T \to \infty} \frac{1}{T} \int_0^{T} dt \langle \hat{\mathcal{O}}(t)\rangle
\end{equation}
for any operator $\hat{\mathcal{O}}$. Fig.~\ref{fig_Chirality}(c) shows the average current as a function of drive strength $\Omega/J$. The chirality vanishes at $\Omega = 0$ (for which the ladder legs decouple and $\hat{I}_c$ commutes with the non-interacting Hamiltonian) and at $\Omega/J \to \pm \infty$ (for which the initial state is an eigenstate of the Hamiltonian), and is strongest at intermediate values of $\Omega / J \approx \pm 2$. For $L \to \infty$ we can write this non-interacting average analytically as,
\footnotesize
\begin{equation}
\label{eq_SU2nonInteractingExact}
    \frac{\langle \hat{I}_{c}\rangle_{\infty}}{L}\bigg|_{U=0} = \frac{\Omega/J}{2} \cot \left(\frac{\phi}{2}\right)\left(1-\frac{|\Omega/J|}{\sqrt{(\Omega/J)^2 + 16 \sin^2 (\phi/2)}}\right).
\end{equation}
\normalsize
The overall direction of the chiral current has odd parity symmetry under $\phi \to - \phi$ or $\Omega \to - \Omega$, while the strength of the current vanishes at $\phi = 0$ (no magnetic flux) or $\phi=\pi$ (left-right reflection symmetry is restored) and tends to be strongest at around $\phi \approx \pi/2$.

We now turn to the strongly interacting regime. In the $U\gg J$ limit, a Fermi-Hubbard model without driving or spin-orbit coupling at unit filling is in the Mott insulator regime, which would suppress particle current because of the high energetic cost of forming a doublon. In the presence of the synthetic gauge field that breaks SU(2) symmetry this behavior is modified. In Fig.~\ref{fig_Chirality}(d) we show a sample time evolution of the current for strong interactions $U/J= 30$, flux $\phi = \pi/2$, and two different drive strengths $\Omega/J =2$ and $\Omega/J = 32$. The weak drive exhibits the expected behaviour where dynamics is suppressed as the system is an insulator. For the strongly driven case, on the contrary we observe a significant current since the interaction energy penalty for generating a doublon is partially compensated by the drive.

Fig.~\ref{fig_Chirality}(e) shows the long-time average current across the range of $\Omega/J$. We find that the strongest response occurs around $\Omega\approx U$ with a profile similar to the non-interacting case, but shifted to $\Omega=U$ instead of $\Omega=0$. A small quench of drive strength from e.g. $\Omega = U - \epsilon$ to $U + \epsilon$ for $\epsilon \sim J$ can thus cause a macroscopic reversal of the current without breaking any obvious reflection symmetry in the Hamiltonian. Furthermore, there are several other points where the current undergoes non-trivial behavior or changes direction altogether. These special points are located at fractional values of $\Omega = U/2, U/3, \dots$ and correspond to multi-body resonances. We show the average current as a function of both $\Omega/U$ and $\phi$ for fixed $U/J$ in Fig.~\ref{fig_Chirality}(f); the resonant features persist across the parameter range, growing stronger or weaker depending on the flux and resonance in question.

A finite flux therefore enables the generation of particle transport and chiral currents in the limit where interactions would normally inhibit motion. Such behavior is best understood in the diagonal frame, which we will explore next.

%%%
\subsection{Resonant dynamics of dressed states}
%%%

From a physical perspective, the rate of particle flow along the legs of the synthetic ladder quantifies its longitudinal conductivity. In addition, the system can have motion along the synthetic dimension. Since the initial state is the dressed spin state $\uparrow$, which is a delocalized superposition of atoms along the synthetic dimension in the lab frame, a change in the $\uparrow$ dressed state population during the dynamics is a measure of the transverse conductivity of the system which we will study using the diagonal frame.

\begin{figure}
\includegraphics[width=.48\textwidth]{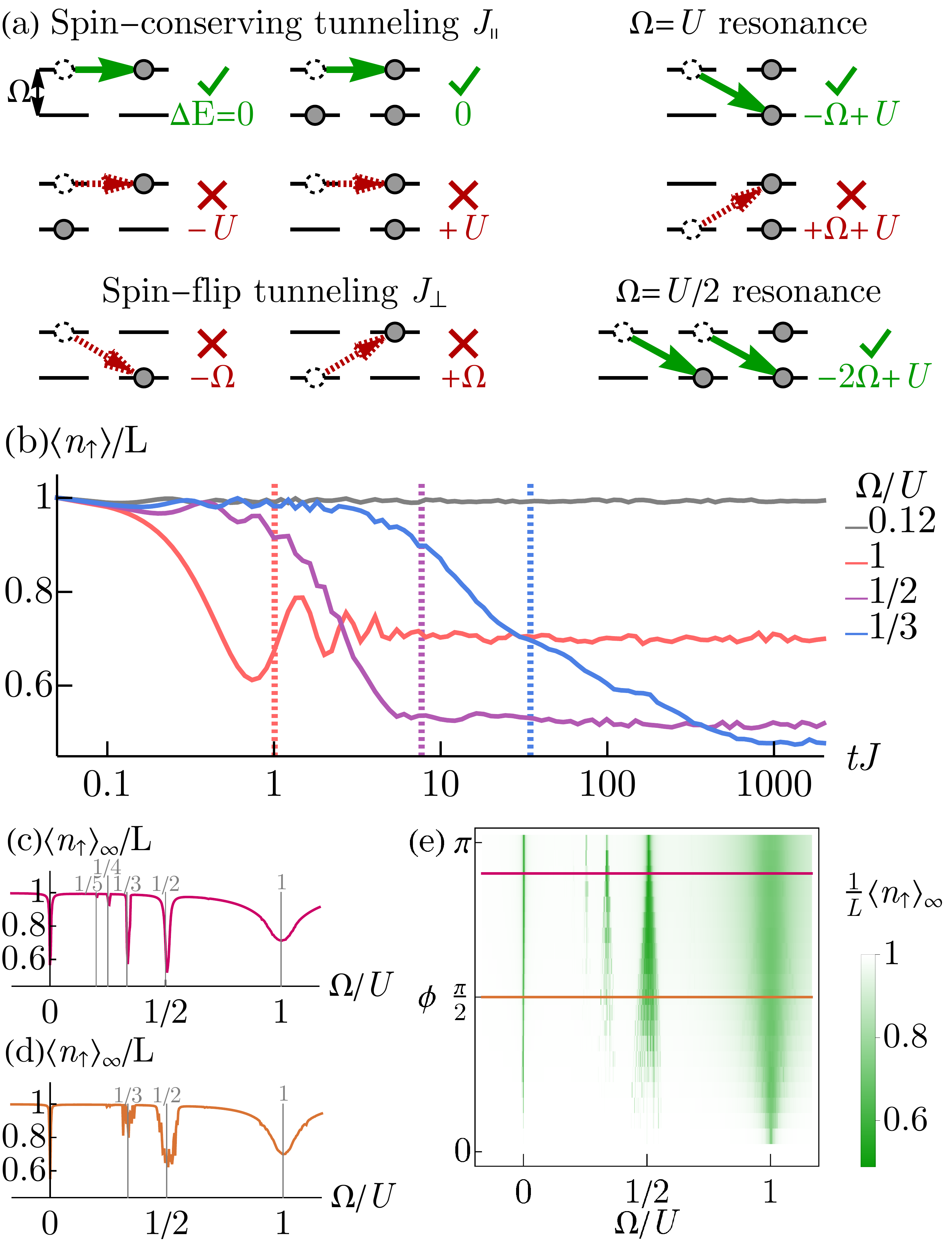}
\centering
\caption{(a) Diagrams of the allowed and forbidden tunneling processes of the $n=2$ system in the diagonal frame for $\Omega, U \gg J$. The energy cost for an atom hopping from one level (indicated by an empty circle) to another (indicated by the arrow) is shown. Processes with green arrows cost zero energy and are allowed, while ones with red arrows are off-resonant and forbidden. Interaction-enabled tunneling processes are shown for the $\Omega = U$ and $\Omega = U/2$ resonances, although higher-order ones can occur as well. (b) Time-evolution of the dressed state population $\langle\hat{n}_{\uparrow}\rangle$ for a quench starting with all atoms in that state, for a system size $L=11$, large flux $\phi = 9\pi/10$, strong interactions $U/J = 30$, and selected drive strengths $\Omega / U = 0.12$ (non-resonant) and $\Omega/U = 1, 1/2, 1/3$ (different resonances). For the latter, the associated timescales $tJ_{\perp}=1$, $4t J_{\perp}^2/U = 1$, $27t J_{\perp}^3/U^2= 1$ are indicated by vertical dotted lines. (c-d) Long-time average of the population $\langle\hat{n}_{\uparrow}\rangle_{\infty}$ with $U/J=30$, system size $L=9$, and flux values of (c) $\phi = 9\pi/10$ and (d) $\phi = \pi/2$, averaging out to time $tJ=500$. Discernable resonances $\Omega = \frac{1}{q}U$ are labeled. (e) Average population across the parameter range of $\Omega/U$ and $\phi$, still using $U/J = 30$ and $L=9$. The colored lines corresponds to the preceding panels (c-d).}
\label{fig_Population}
\end{figure}

For generic flux the dressed spin atoms can undergo both spin-conserving tunneling with rate $J_{\parallel}$, and spin-flipping tunneling with rate $J_{\perp}$. However, the actual motion is subject to kinetic constraints enforced by energy conservation. We consider the regime of $\Omega, U \gg J$ and $\Omega \sim U$ where the interactions and drive set the dominant energy scale, and the non-trivial resonant features occur. The resulting constraints are summarized in Fig.~\ref{fig_Population}(a). In general, fermions are Pauli blocked from tunneling into occupied states. Furthermore, spin-conserving tunneling set by $J_{\parallel}$ is only allowed if no doublon is created or destroyed, as otherwise it would incur a Hubbard interaction penalty $\pm U$. Spin-flipping tunneling $J_{\perp}$ of a lone atom into an empty site incurs an energy penalty $\pm\Omega$ from the drive and is also inhibited. However, depending on the ratio of $\Omega/ U$ there can be a spin-flipping process that is enabled via resonance.

The simplest example is $\Omega = U$. If an $\uparrow$ atom tunnels into $\downarrow$ on a site that already has an $\uparrow$ atom of its own, forming a doublon, the total energy cost is $- \Omega + U = 0$ as shown in Fig.~\ref{fig_Population}(a). This interaction-enabled tunneling process may occur freely, and two adjacent $\uparrow$ atoms can propagate in a a leap-frog like manner~\cite{mamaevLeapFrogs2019}. An analogous process of two neighbouring $\downarrow$ atoms forming a doublon cannot occur since the associated cost is $+\Omega + U \neq 0$. This resonance causes the qualitative shift of the current profiles from $\Omega = 0$ to $\Omega = U$ [Figs.~\ref{fig_Chirality}(b) and (c)]. At the resonance the dynamics are similar to those of free fermions, up to additional tunneling constraints since pairs of adjacent $\downarrow$ atoms cannot leap-frog move.

When one is instead close to the resonance but not exactly at it, there are high-order chiral tunneling processes in perturbation theory proportional to powers of $1/(U-\Omega)$ that change sign as the resonance is passed (for odd powers). This is the cause of the current direction reversal across the resonance.

Other resonances manifest through higher-order tunneling processes. For $\Omega = U/2$, there is a second order process where two adjacent $\uparrow$ atoms both tunnel and flip to $\downarrow$, forming one doublon using a third $\uparrow$ atom, as shown in Fig.~\ref{fig_Population}(a). The total cost of this process is $-2\Omega + U = 0$. The overall rate can be estimated using degenerate perturbation theory as $\frac{J_{\perp}}{-\Omega + U}J_{\perp} \times 2 = \frac{4J_{\perp}^2}{U}$, where the denominator is the energy of the intermediate state of the process and the $\times 2$ comes from the number of ways in which the two steps can be taken. Similarly at $\Omega = U/3$ we have a three-step resonance with approximate rate $\frac{J_{\perp}}{-\Omega + U}\frac{J_{\perp}}{-\Omega}J_{\perp} \times 6 = \frac{27J_{\perp}^3}{U^2}$ (with a factor of $\times 6$ since a three-step process can be ordered in 6 possible ways). In general we can predict resonances of the form $\Omega = \frac{p}{q} U$ for integer $p$, $q$, each corresponding to a $(p + q)$-atom process that moves $q$ atoms down from $\uparrow$ to $\downarrow$ and forms $p$ doublons. Resonances with $p > 1$ correspond to correlated processes where multiple doublons are formed in different, possibly distant parts of the lattice. Such resonances are less prominent, but can still manifest as non-monotonic features in the average dynamics if the system is evolved to sufficiently long timescales. Although they have a different physical origin, in many ways these resonances resemble the ones observed in the longitudinal resistance in the fractional quantum Hall effect, at fractional filling fractions as the magnetic field is changed.

We characterize these resonances by measuring the population of the initially filled dressed state $\langle \hat{n}_{\uparrow}\rangle = \sum_{j} \langle\hat{n}_{j,\uparrow} \rangle$. When far from any resonances, for strong $\Omega \gg J$ the initial state is disparate in energy from the rest of the spectrum. All tunneling is thus inhibited and no dynamics will occur, at least up to timescales that grow exponentially long with system size~\cite{Sensarma2011}. At a resonance, interaction-enabled tunneling will cause a reduction of the population over a timescale set by the corresponding rate, which is $J_{\perp}$ for the $\Omega=U$ resonance, $4J_{\perp}^2/U$ for the $\Omega = U/2$ resonance, $27J_{\perp}^3/U^2$ for $\Omega = U/3$, etc..

Fig~\ref{fig_Population}(b) shows characteristic time evolution profiles of this population. In the long time limit of the relevant rate we see saturation to an average value $\langle\hat{n}_{\uparrow} \rangle_{\infty}$. We numerically compute this average for different $\Omega/U$, $\phi$ at fixed $U/J$ in Figs.~\ref{fig_Population}(c-d). Increasing $\phi$ makes the resonances stronger and permits the resolution of higher-order ones because the spin-flip tunneling $J_{\perp}$ grows with $\phi$. There is also a resonance at $\Omega = 0$ that corresponds to conventional antiferromagnetic superexchange interactions; while there is no drive in that regime, the initial state is a spiral state in the lab frame with both $e$ and $g$ atom population which still exhibits dynamics.

All of these features are not exclusive to our specific quench; we have just chosen an initial state for which they are particularly prominent. For example, one could consider a different initial state such as $\prod_j \hat{c}_{j,e}^{\dagger}\ket{0} = \prod_j \frac{1}{\sqrt{2}}\left(\hat{a}_{j,\downarrow}^{\dagger} + \hat{a}_{j,\uparrow}^{\dagger}\right)\ket{0}$, which corresponds to all atoms in the lab frame spin state $e$ sitting in one leg of the ladder. Such a state is a superposition of Fock states with different $\uparrow$, $\downarrow$ configurations in the diagonal frame, some of which will have energetically allowed tunneling processes and will undergo dynamics (see Appendix~\ref{app_DifferentInitial} for details).

%%%
\subsection{Relaxation}
%%%

\begin{figure}
\includegraphics[width=.48\textwidth]{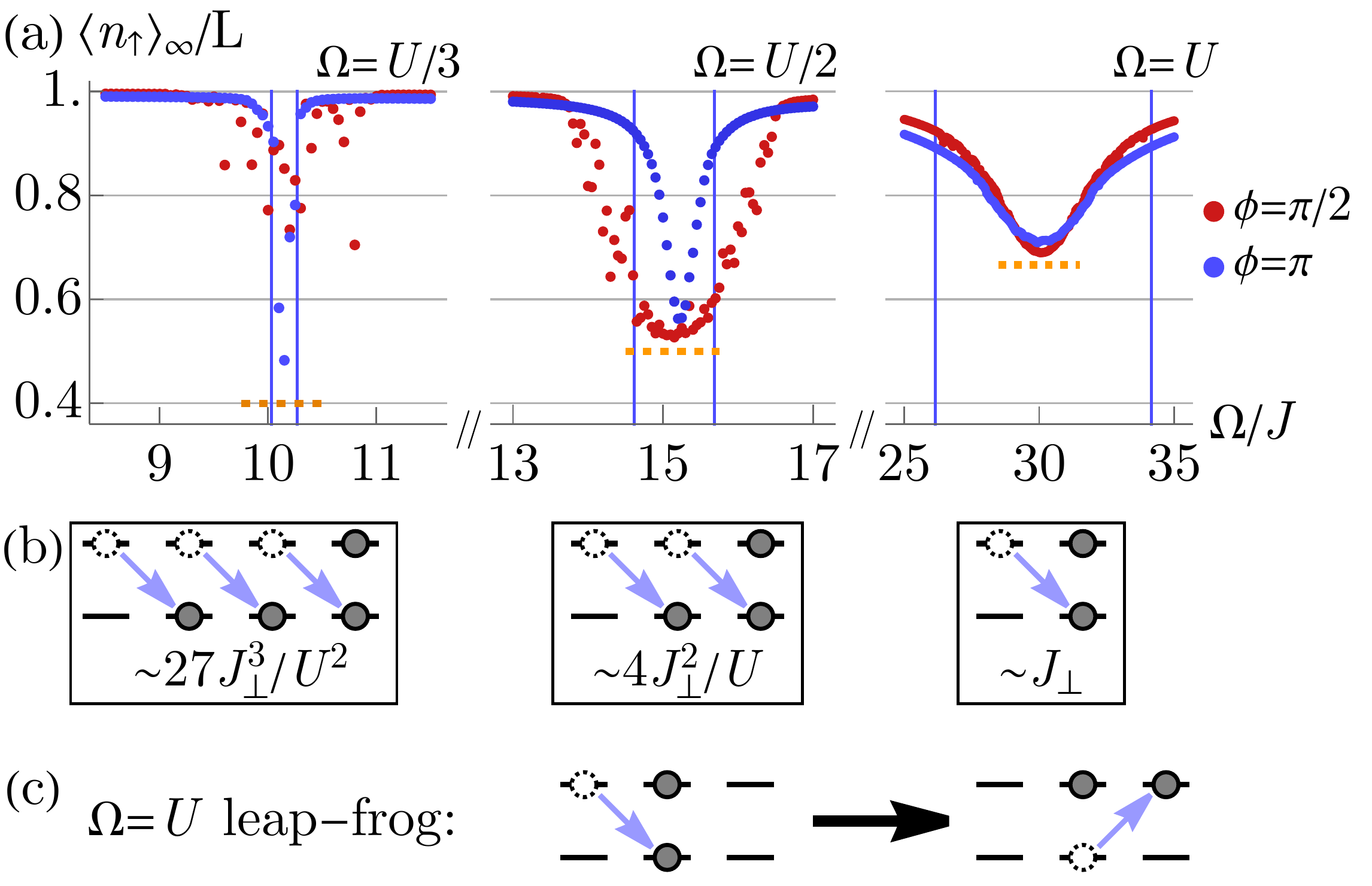}
\centering
\caption{(a) Long-time average of the population $\langle \hat{n}_{\uparrow}\rangle$ near the resonances $\Omega = U/3, U/2, U$, for different flux $\phi = \pi/2, \pi$. The system size is $L=11$ and the interaction strength is $U/J = 30$. Time evolution is done out to $tJ = 1500$, and averaging is done over $tJ = 750$ to $1500$ to avoid the initial population decay. The $\phi = \pi$ resonance widths are set by the respective rates $\frac{27J_{\perp}^3}{U^2}$, $\frac{4J_{\perp}^2}{U}$, $J_{\perp}$; the vertical blue lines represent values of $\pm 4$ times the corresponding rate for each resonance. The $\phi = \pi/2$ resonance widths are instead broadened by an effective bandwidth $\sim J_{\parallel}$. Note that the specific positions of the resonances are also shifted due to off-resonant super-exchange interactions by a factor on the order of $\sim J^2/U$. The orange dashed lines correspond to infinite-temperature thermal average predictions of $2/5$, $1/2$ and $2/3$ respectively (see Appendix~\ref{app_Thermal}). (b) Associated interaction-enabled tunneling processes for each respective resonance in (a). (c) Leap-frog process by which pairs of $\uparrow$ atoms can move at the $\Omega = U$ resonance via $J_{\perp}$ only.}
\label{fig_SU2Thermalization}
\end{figure}

The behavior of the long-time average dressed state population is determined by the system's ability to equilibrate via tunneling. At resonance the interaction-enabled tunneling processes allow the system to relax by exploring the Hilbert space that is energetically accessible. We plot the average population close to various resonances in Fig.~\ref{fig_SU2Thermalization}(a) for different flux values $\phi = \pi/2$ and $\phi = \pi$. The results for the $\Omega=U/3$ resonance appear irregular due to finite-size and finite-time effects, but are expected to become smooth in the limit of $L \to \infty$; see Appendix~\ref{app_Scaling} for details.

For $\phi = \pi$, the system can only have spin-flip tunneling $J_{\perp}= J$ while $J_{\parallel} = 0$. Each resonance has one allowed tunneling process under this constraint. For $\Omega = U/q$ with $q \in \{1,2,3,\dots\}$ we flip $q$ atoms from $\uparrow$ to $\downarrow$ and form one doublon. This process is depicted for each resonance in Fig.~\ref{fig_SU2Thermalization}(b) for clarity. The many-body dynamics amount to the coherent superposition of this process occurring in parallel across the lattice. The widths of the resonances are set by the rate of the process, which is $J_{\perp}$ for $\Omega=U$, $4J_{\perp}^2/U$ for $\Omega = U/2$, $27J_{\perp}^3/U^2$ for $\Omega = U/3$. These widths are indicated by the blue dashed lines in Fig.~\ref{fig_SU2Thermalization}(a) and show good agreement with the $\phi = \pi$ results.

The above discussion is valid at short times because the initial state is unit-filled and any spin-conserving tunneling is Pauli-blocked. The interaction-enabled spin-flip tunneling process is the only one that can occur at time $t=0$ even if $J_{\parallel} \neq 0$. However, once interaction-enabled tunneling does occur, it leaves behind a hole. States with a hole can be coupled to each other via spin-conserving tunneling processes, resulting in a dispersion that broadens the otherwise degenerate spectrum of the energetically accessible subspace and thus the resonance. For the $\Omega = U$ resonance, atoms can freely move into a hole using either spin-conserving tunneling $J_{\parallel}$ directly or interaction-enabled spin-flip tunneling $J_{\perp}$ via the leap-frog process shown in Fig.~\ref{fig_SU2Thermalization}(c). The resonance width is then determined by both $J_{\parallel}$ and $J_{\perp}$, which are comparable for both studied flux values. Higher-order resonances require a finite $J_{\parallel}$ for atoms to move into a hole. The matrix element $J_{\parallel}$ is non-vanishing at $\phi = \pi/2$, and broadens the associated resonances for that flux by an effective bandwidth $\sim J_{\parallel}$. The associated timescale for relaxation correspondingly increases as the resonance broadens, since the resonant tunneling must overcome the splitting of the many-body states induced by the dispersion.

The height of the resonance peaks can be predicted with a thermal average. At resonance the system has a set of energetically accessible many-body states. Since these states all have the same energy as the initial state (neglecting broadening effects), the thermal description is given by a microcanonical ensemble with an effectively infinite temperature. The corresponding thermal state is written as $\rho_{\mathrm{th}} = \hat{P}_{\mathrm{res}}/n_{\mathrm{res}}$ with $\hat{P}_{\mathrm{res}}$ the projector onto the accessible subspace and $n_{\mathrm{res}}$ the size of the subspace. The prediction for the observable is then $\langle \hat{n}_{\uparrow}\rangle_{\infty} = \text{tr}(\hat{n}_{\uparrow}\rho_{\mathrm{th}}) = \text{tr}(\hat{P}_{\mathrm{res}}\hat{n}_{\uparrow})/n_{\mathrm{res}}$. In the limit $L \to \infty$ this predicted value is $\langle \hat{n}_{\uparrow}\rangle_{\infty} = \frac{2}{3}, \frac{1}{2}, \frac{2}{5}$ for the $\Omega=U, U/2, U/3$ resonances respectively (see Appendix~\ref{app_Thermal} for derivation). The peak resonance values in Fig.~\ref{fig_SU2Thermalization}(a) show reasonable agreement with these thermal predictions, aside from the $\Omega = U/3$ resonance which would likely require longer timescales and larger systems due to its high-order nature. A numerical analysis on convergence is given in Appendix~\ref{app_Scaling}.

We note as a caveat that in general, kinetic constraints upon the motion can inhibit the ability of the system to explore even the resonantly accessible Hilbert space and prevent agreement with a thermal average~\cite{lan2018relaxation}, although we see reasonable agreement for the resonances studied here. In addition, for large but not infinite $U/J$, $\Omega/J$ (as in a real experimental context) the wavefunction is not guaranteed to be fully confined to the accessible Hilbert space. For large enough systems some of the energy carried by the doublons and drive excitations can be dispersed into kinetic energy, causing leakage into other Hilbert space sectors, although the system's capacity for this depends on the initial state and structure of the many-body spectrum. A full characterization of the system's infinite-time thermodynamic limit properties encoding the above complications is an interesting pursuit, albeit one beyond the scope of the current work.

%%%%%
\section{SU($n$) system}
\label{sec_SUn}
%%%%%
%%%
\subsection{Fermi-Hubbard model}
\begin{figure*}
\includegraphics[width=1\textwidth]{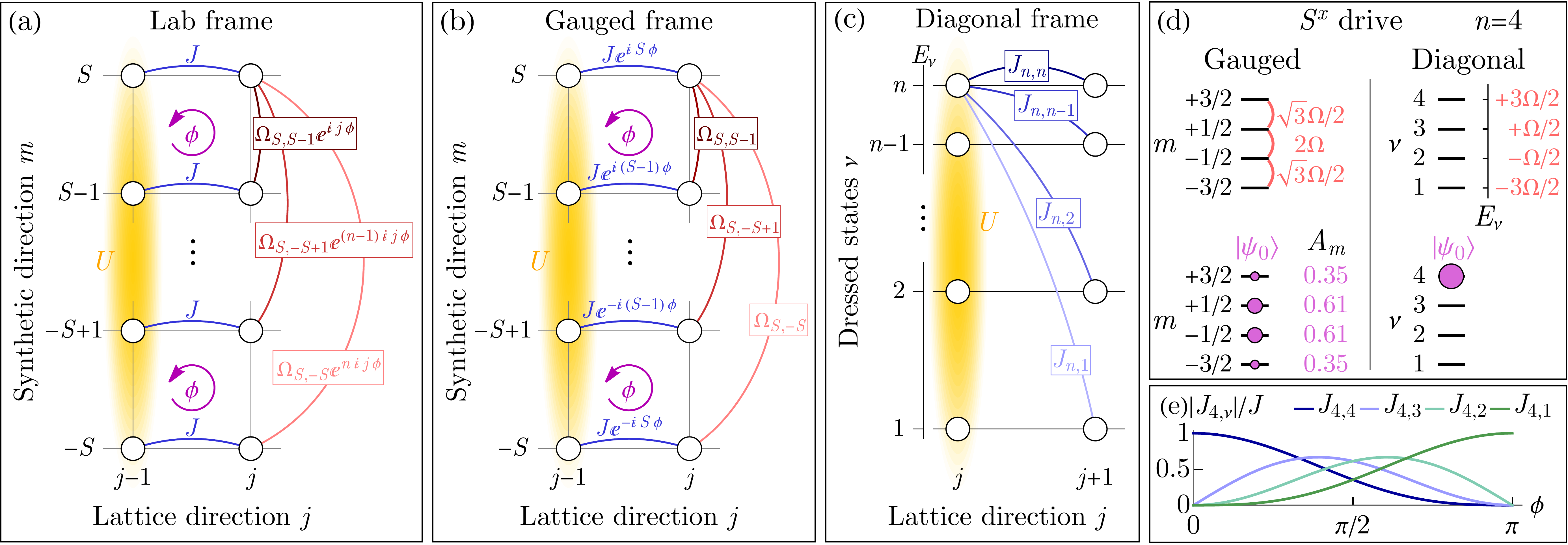}
\centering
\caption{(a) Schematic of the $n>2$ system now engineering an $n$-leg ladder in the lab frame. The synthetic dimension states are indexed $m \in \{-S,-S+1,\dots,S\}$ with $S=(n-1)/2$ the size of an effective on-site spin. Motion along the synthetic direction is induced by a laser with couplings $\Omega_{m,m'}$ between internal states $m$, $m'$ and a phase $e^{i j (m-m')\phi}$ to emulate the magnetic field. Only the couplings from the $m=S$ state to the others are shown for simplicity, although generically each spin state can be coupled to each other. The SU($n$) symmetric interactions are all-to-all in the synthetic dimension and on-site in the lattice dimension. (b) Visualization of the same system in the gauged frame where the flux is put into the tunneling terms. (c) Visualization of the system in the diagonal frame, where the synthetic spin dimension is comprised of dressed spin states $\nu \in \{1, \dots, n\}$ that are single-particle eigenstates of the drive. In this frame the nearest-neighbour tunneling can flip spin from $\nu$ to $\nu'$ with corresponding matrix element $J_{\nu,\nu'}$. Again, only couplings from $\nu = n$ are shown for simplicity although all dressed states generically couple to all others. (d) Couplings between spin states in the gauged frame and corresponding dressed energy levels in the diagonal frame for an $\hat{S}^{x}$ type drive and $n=4$ states. The typical initial state of all atoms in the highest energy dressed state $\nu = 4$ is shown in both frames with the corresponding wavefunction amplitude $A_m$ of each spin state in the gauged frame. (e) Tunneling matrix elements $J_{\nu,\nu'}$ allowed for the initially occupied dressed state $\nu = 4$ for the drive in panel (d).}
\label{fig_SUnSchematic}
\end{figure*}

%%%
We now move to the situation where the laser drive couples a larger number of $n > 2$ internal levels. The Hamiltonian maintains the same form as Eq.~\ref{eq_FermiHubbardHamiltonian}, only now involving $n$ spin flavors labeled by $m \in \{-S, -S+1, \dots, S\}$ with $S = (n-1)/2$ the size of the effective on-site spin. Such a system emulates a synthetic 2D strip of size $L \times n$. If all the internal levels experience the same lattice potential, the tunneling Hamiltonian takes the form,
\begin{equation}
\hat{H}_{\mathrm{J}} = -J \sum_{j,m} \left(\hat{c}_{j,m}^{\dagger}\hat{c}_{j+1,m}+h.c.\right).
\end{equation}
For the case of SU($n$) symmetric interactions, collisions between any pair of internal states are characterized by the same scattering length~\cite{Cazalilla2014, reyTwoOrbitalSUNMagnetism2010,cazalilla2009sunPhases} and the on-site Hubbard repulsion can be written as,
\begin{equation}
\hat{H}_{\mathrm{U}} = \frac{U}{2}\sum_{j,m,m'} \hat{n}_{j,m}\left(\hat{n}_{j,m'}-\frac{1}{n}\mathbbm{1}\right),
\end{equation}
where $\mathbbm{1}$ is the identity operator. When visualizing the internal levels as a synthetic ladder, the interactions become all-to-all along the synthetic direction and on-site along the lattice direction. The laser drive coupling the levels along the synthetic direction can be written in general as,
\begin{equation}
\hat{H}_{\Omega} = \sum_{j,m,m'}\Omega_{m,m'} e^{i j (m-m') \phi} \hat{c}_{j,m}^{\dagger}\hat{c}_{j,m'}.
\end{equation}
The coefficient $\Omega_{m,m'}$ gives the strength of the coupling matrix element between spin states $m$ and $m'$. The phase $e^{i j (m-m')\phi}$ ensures that every plaquette of the synthetic lattice is pierced by a constant flux $\phi$ (see Section~\ref{sec_Implementation} and Ref.~\cite{perlinSUn2021} for details on experimental implementation). Fig.~\ref{fig_SUnSchematic}(a) shows this system in the lab frame. Note that experimentally realistic implementations will generally be constrained to couple spin states with $|m - m'| \leq 2$, because every change of $m$ by one requires an additional photon and laser coupling schemes tend to be two-photon at most. We keep generic laser couplings for now, and will provide specific realistic examples further on. 

As in the $n=2$ case, we make a gauge transformation to put the flux into the tunneling terms instead of the drive,
\begin{equation}
\hat{c}_{j,m} \to e^{i j m \phi} \hat{c}'_{j,m},
\end{equation}
In this gauge the tunneling becomes,
\begin{equation}
\hat{H}_{\mathrm{J}} = -J\sum_{j,m} \left(e^{i m \phi}\hat{c}_{j,m}'^{\dagger}\hat{c}'_{j+1,m} +h.c.\right),
\end{equation}
while the drive becomes,
\begin{equation}
\hat{H}_{\Omega} = \sum_{j,m,m'} \left(\Omega_{m,m'}\hat{c}_{j,m}'^{\dagger}\hat{c}'_{j,m'}+h.c.\right).
\end{equation}
The interactions remain unchanged. The resulting system in the gauged frame is depicted in Fig.~\ref{fig_SUnSchematic}(b).

For a fixed $n$ we can again make a single-particle basis transformation to diagonalize the drive. If we treat the spin states as first-quantized states of angular momentum projection $\ket{m}$, this transformation can be written as,
\begin{equation}
\hat{W}^{\dagger} \left(\sum_{m,m'}\Omega_{m,m'}\ket{m}\bra{m'}\right)\hat{W} = \sum_{\nu} E_{\nu} \ket{\nu}\bra{\nu},
\end{equation}
where $\hat{W} = \sum_{m,m'} W_{m,m'}\ket{m}\bra{m'}$ is the corresponding unitary and $\nu \in \{1, \dots, n\}$ indexes over the drive eigenstates with eigenenergies $E_{\nu}$. As a simple example to connect with the previous section, for the $n=2$ case we have:
\begin{equation}
\begin{aligned}
    n=2:\>\>\>&\sum_{m,m'}\Omega_{m,m'}\ket{m}\bra{m'} = \frac{\Omega}{2} \left(\begin{array}{cc}0 & 1\\ 1& 0\end{array}\right),\\
    &\hat{W} = \frac{1}{\sqrt{2}} \left(\begin{array}{cc}1 & -1\\ 1 & 1 \end{array}\right),
\end{aligned}
\end{equation}
in the basis of $\{\ket{+\frac{1}{2}}, \ket{-\frac{1}{2}}\}$ (equivalent to $\{e,g\}$).

Using this transformation, we define new dressed spin operators as
\begin{equation}
\hat{a}_{j,\nu} = \sum_{m} W_{m,\nu-S-1}^{*} \hat{c}'_{j,m}.
\end{equation}
The resulting drive in the diagonal frame becomes,
\begin{equation}
\hat{H}_{\Omega} = \sum_{j,\nu} E_{\nu}\hat{n}_{j,\nu}.
\end{equation}
The interactions maintain SU($n$) symmetry and keep the form $\hat{H}_{\mathrm{U}} = \frac{U}{2}\sum_{j,\nu,\nu'}\hat{n}_{j,\nu}\left(\hat{n}_{j,\nu'}-\frac{1}{n}\mathbbm{1}\right)$. The tunneling now contains generalized spin-flip terms,
\begin{equation}
\label{eq_DressedTunnelingRate}
\begin{aligned}
\hat{H}_{\mathrm{J}} &= \sum_{j,\nu,\nu'} \left(J_{\nu,\nu'}\hat{a}_{j,\nu}^{\dagger}\hat{a}_{j+1,\nu'} + h.c.\right),\\
J_{\nu,\nu'} &= -J\sum_{m} e^{i m \phi} W_{m,\nu-S-1}^{*} W_{m,\nu'-S-1},
\end{aligned}
\end{equation}
with $J_{\nu,\nu'}$ the matrix element for an atom to tunnel one lattice site over and flip from dressed spin $\nu$ to $\nu'$. When $\phi = 0$, we only have spin-conserving tunneling $J_{\nu,\nu'} = -J \delta_{\nu,\nu'}$. For generic $\phi$, the strength of the various tunneling amplitudes will depend on both $\phi$ and the structure of the laser couplings $\Omega_{m,m'}$. Fig.~\ref{fig_SUnSchematic}(c) shows the system in the diagonal frame. This spin-flipping $\phi$-dependent tunneling has been described in the literature as a ``flavor-orbital'' coupling~\cite{ghoshAttractiveFermiGasesBaryonSUN2017}. In our framework, having a generic $\phi \neq 0, \pi$ greatly enhances the connectivity of the system because any dressed spin can turn into any other dressed spin, thus enabling a much larger set of multi-body resonances.

While we wrote arbitrary coefficients $\Omega_{m,m'}$, realistic schemes will be more constrained. We will explore a scheme that gives clear insight on how the $n>2$ systems behave in the strongly interacting limit. Specifically, we will consider couplings of the form $\Omega_{m,m+1} = \Omega_{m+1,m} = \frac{\Omega}{2}\sqrt{(S-m)(S+m+1)}$ and zero otherwise, with $\Omega$ an overall drive strength. This configuration yields a drive that looks like a spin-$S$ operator $\hat{S}^{x}$ in the gauged frame, if one treats the internal states as Zeeman levels of a spin-$S$ particle. The unitary transformation  $\hat{W}$ can be written as $\hat{W} = e^{-i \frac{\pi}{2}\hat{S}^{y}}$, which diagonalizes the drive by rotating it into $\hat{W}^{\dagger} \hat{S}^{x} \hat{W}=\hat{S}^{z}$. The resulting drive eigenenergies will be equally spaced by $\Omega$, with $E_{\nu} = \Omega (\nu-S-1)$ if we label $\nu = 1 \dots n$. In Fig.~\ref{fig_SUnSchematic}(d) we show the couplings and corresponding dressed state energies for an $n=4$ system. The tunneling rates of the dressed states in the diagonal frame are shown in Fig.~\ref{fig_SUnSchematic}(e).

After studying the behavior of such a configuration, we will discuss how to experimentally implement it, as well as other more generic cases. Alternative distributions of the coupling, such as uniform couplings with periodic boundary conditions, have also been theoretically~\cite{cooper2015periodicSynthetic} and experimentally~\cite{hanToblerone2019,Liang2021} studied.

%%%
\subsection{Chirality}
%%%

\begin{figure*}
\includegraphics[width=1\textwidth]{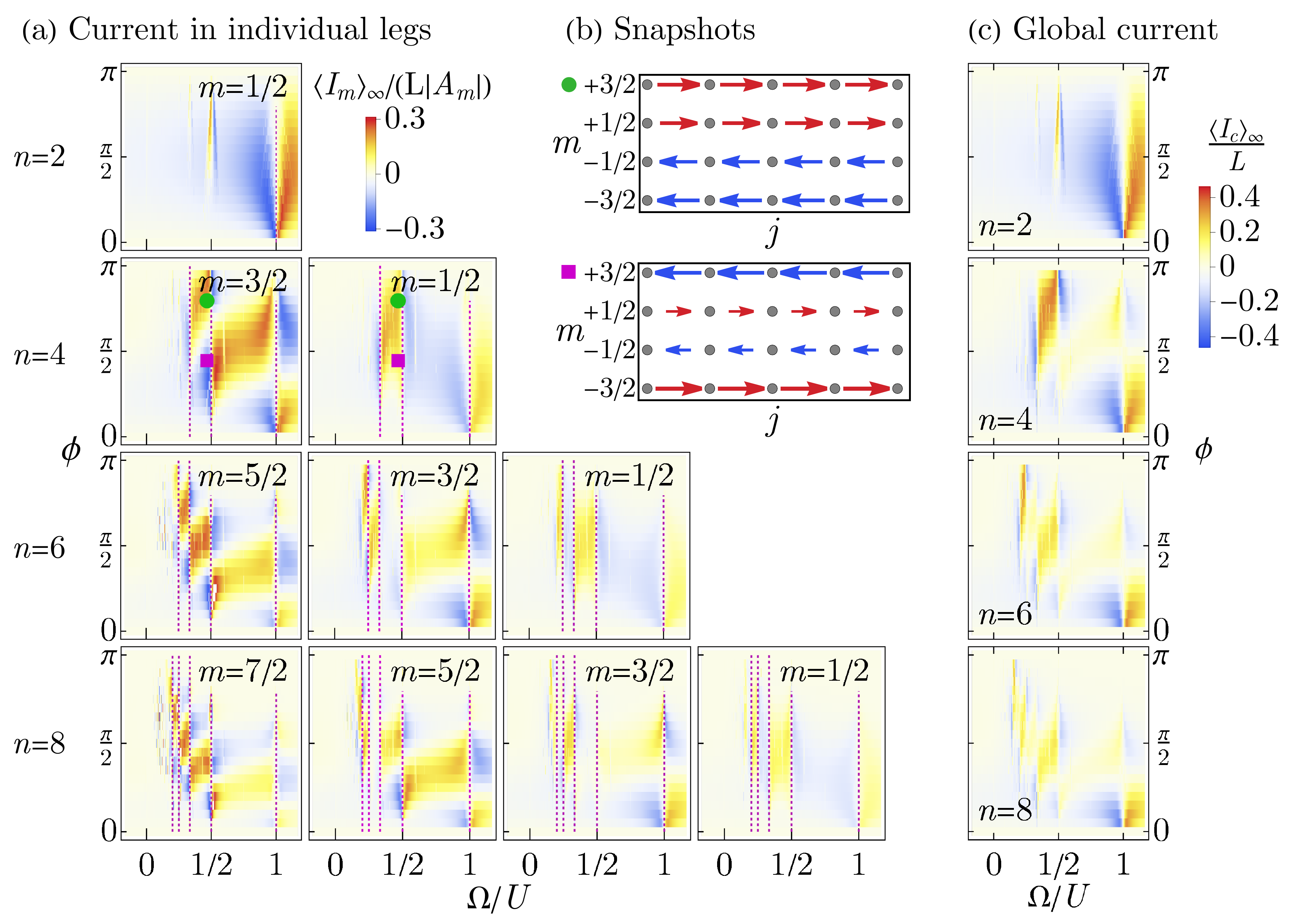}
\centering
\caption{(a) Average spin current in each leg $m$ for different synthetic dimension sizes $n$ across the parameter range of $\Omega/U$ and $\phi$. We scale the current along leg $m$ by the initial wavefunction probability amplitude $|A_m| = |\langle \psi_0 | m \rangle|$ in that leg. Prominent resonant crossovers are highlighted with dashed lines. Only $m > 0$ legs are shown; the other legs are equal and opposite, $\langle\hat{I}_{m}\rangle = -\langle \hat{I}_{-m}\rangle$. The system size is $L=9, 5, 4, 4$ for $n=2,4,6,8$ respectively. Interactions are fixed at $U/J = 30$. The system is evolved to time $tJ = 500$ before taking the average. (b) Snapshots of current profiles for the $n=4$ system in two particular regions of parameter space indicated by the green and purple markers in panel (a). The size of the arrows is proportional to the scaled average current along that link; current across the periodic link is not shown. (c) Overall average chiral current obtained by summing over all legs (with no scaling factors) for the same parameters as (a). }
\label{fig_SUnChirality}
\end{figure*}

\begin{figure}
\includegraphics[width=0.48\textwidth]{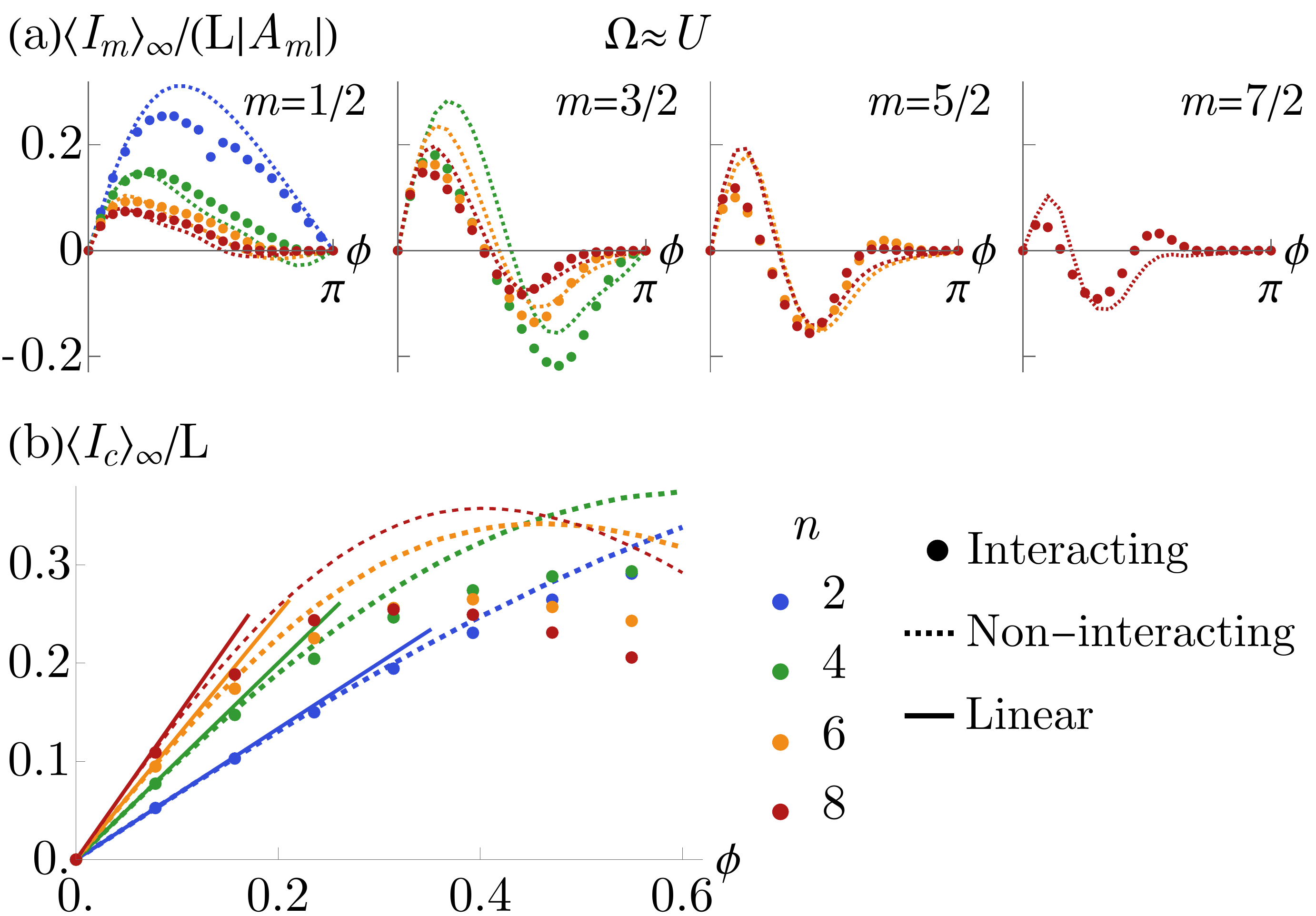}
\centering
\caption{(a) Average scaled current in different legs $m$ close to the $\Omega = U$ resonance for $n=2,4,6,8$, fixing $U/J = 30$ and $\Omega/J = 33$. The dots show the numerically computed data for the strongly interacting system, sizes $L = 9,5,4,4$ as before, evolved to time $tJ = 500$. The dashed line is the corresponding average for a non-interacting system $U=0$ with the same initial condition and effective drive strength $\Omega / J = 3$, the same distance from the $\Omega = U$ resonance assuming that it is shifted from $\Omega = U$ to $\Omega = 0$. (b) Average chiral current for the same resonance in the $\phi \ll 1$ limit. The solid lines are a linear approximation with a slope given by Eq.~\eqref{eq_AnalyticSlope}.}
\label{fig_SUnChiralityNonInt}
\end{figure}

As in the $n=2$ case we study the quench dynamics of spin currents for a unit filled system $N/L = 1$ using exact numerical time-evolution. We initialize a state of all atoms in the highest-energy drive eigenstate,
\begin{equation}
\ket{\psi_0} = \prod_{j}\hat{a}^{\dagger}_{j,\nu=n} \ket{0}.
\end{equation}
This initial state is delocalized along the synthetic ladder in the gauged frame and exhibits enhanced resonant features in the dynamics. 

We define the longitudinal current along each leg $m$ of the ladder (in units of $J$),
\begin{equation}
    \hat{I}_{m} = - i \sum_{j} \left( \hat{c}_{j,m}^{\dagger}\hat{c}_{j+1,m} - h.c.\right).
\end{equation}
We again look at the long-time averages $\langle \hat{I}_{m}\rangle_{\infty}$. The currents are scaled by the wavefunction amplitudes $|A_m| = |\bra{m} \nu=n \rangle|$ of the initial state in each leg [see example in Fig.~\ref{fig_SUnSchematic}(d)] for visual clarity. We consider even $n$ since this allows for more natural implementation using fermionic atoms. Since the initial state is symmetric about $m \to -m$ and the system preserves this symmetry, we have $\langle \hat{I}_{m}\rangle = - \langle \hat{I}_{-m}\rangle$ and can thus focus on $m > 0$.

Fig.~\ref{fig_SUnChirality}(a) shows the average currents in each leg for a range of $n$. As $n$ increases we observe a larger number of macroscopic spin flow reversals, corresponding to an increasing number of multi-body resonances. The flow becomes concentrated at the edges of the ladder, although the resonant features persist in all legs. We also notice that the system can settle into different chiral current patterns depending on the sign of the current in each leg. Fig.~\ref{fig_SUnChirality}(b) shows snapshots of the average current in two such regimes for $n=4$. The flow can either separate into staggered $L \times 2$ blocks when the $m = 1/2, 3/2$ legs have opposite sign of current, or generate a bulk flow if the sign is the same for both. Higher values of $n$ can generate even more complex patterns.

We can gain more intuition by considering the generalization of the shearing current studied in Section~\ref{subsec_SU2Chirality}. We define the overall average chiral current as (still assuming $n$ even),
\begin{equation}
    \langle \hat{I}_{c}\rangle = \sum_{m >0} \langle \hat{I}_{m}\rangle
- \sum_{m < 0} \langle \hat{I}_{m}\rangle,
\end{equation}
which is shown in Fig.~\ref{fig_SUnChirality}(c).
We observe that as $n$ increases, the overall profile of the current develops a universal shape aside from the emergence of additional resonances at smaller $\Omega/U$. For instance, near $\Omega/U = 1$ and $\phi \ll 1$ the current profiles look very similar for all $n$ up to overall prefactors, suggesting a unifying principle that emerges near that resonance for small flux.

To study this possibility further we look at the currents in each leg close to $\Omega = U$. Recall that for $n=2$, the overall profile of the average chiral current was close to the non-interacting case, only shifted from $\Omega = 0$ to $\Omega = U$. We anticipate that similar behavior will occur for generic $n$. In Fig.~\ref{fig_SUnChiralityNonInt}(a) we compare the average currents near the $\Omega = U$ resonance to a non-interacting result shifted away from the resonance by the same amount, i.e. we compare a non-interacting system with $\Omega /J = 3$ to a strongly interacting one with $(\Omega-U)/J = 3$. We see that for legs deeper inside the bulk of the strip (smaller $|m|$), the average current of the strongly interacting system is well-captured by the equivalent non-interacting system as $n$ increases, especially for smaller flux $\phi$. For this particular resonance at $\Omega = U$, there are sufficiently many interaction-enabled tunneling processes that the dynamics resembles free fermions (at least for the observable in question), as we will further explain in Sec.~\ref{Many}. General closed form analytic solutions for the non-interacting system such as the one in Eq.~\eqref{eq_SU2nonInteractingExact} are not as easy to obtain for $n>2$ as they require the diagonalization of an $n \times n$ matrix. Still, numerical solutions for large $L$ can be straightforwardly computed, which offer insights on the scaling behavior of the system in the $L \to \infty$ and/or $n \to \infty$ limits.

We can also make analytic predictions in certain limiting regimes. As an example, Fig.~\ref{fig_SUnChiralityNonInt}(b) shows the overall average chiral currents $\langle \hat{I}_c\rangle_{\infty}$ of the non-interacting and interacting systems near $\Omega = U$ for the same parameters as above. The two profiles match well as $\phi \to 0$. The non-interacting problem can be analytically solved in the limit of $\phi \ll \frac{J}{\Omega}$, yielding:
\begin{equation}
\label{eq_AnalyticSlope}
    \frac{\langle \hat{I}_{c}\rangle_{\infty}}{L} |_{U = 0} = \frac{n \times n!}{2^{n-1}[(n/2)!]^2}\frac{J}{\Omega}\phi + \mathcal{O}\left[\left(\frac{J}{\Omega}\phi\right)^2\right].
\end{equation}
This Taylor expansion is valid for a non-negligible drive coupling $\Omega$ compared to the lattice tunneling rate $J$. From this expression we obtain a prediction for the interacting system by simply replacing $\Omega$ with $\Omega - U$:
\begin{equation}
    \frac{\langle \hat{I}_{c}\rangle_{\infty}}{L} |_{U \gg J} = \frac{n \times n!}{2^{n-1}[(n/2)!]^2}\frac{J}{\Omega-U}\phi + \mathcal{O}\left[\left(\frac{J}{\Omega-U}\phi\right)^2\right].
\end{equation}
The resulting slope with respect to $\phi$ can be interpreted as the linear response of the system to magnetic fields, as $\phi$ takes the role of a magnetic flux. The small flux limit can be viewed as the continuum limit where the size of the lattice spacing becomes infinitely small. This is a rare case of a strongly interacting system with non-trivial long-time dynamics that can be described via a simple analytic expression.

If we consider a different resonance such as $\Omega = U/2$, a comparison to the non-interacting system no longer suffices because more complex constraints upon the tunneling dynamics emerge. We can gain a better understanding of such a resonance by again turning to the diagonal frame and examining the dressed state population dynamics.

%%%
\subsection{Many-body resonant dynamics}
\label{Many}
%%%

\begin{figure}
\includegraphics[width=0.48\textwidth]{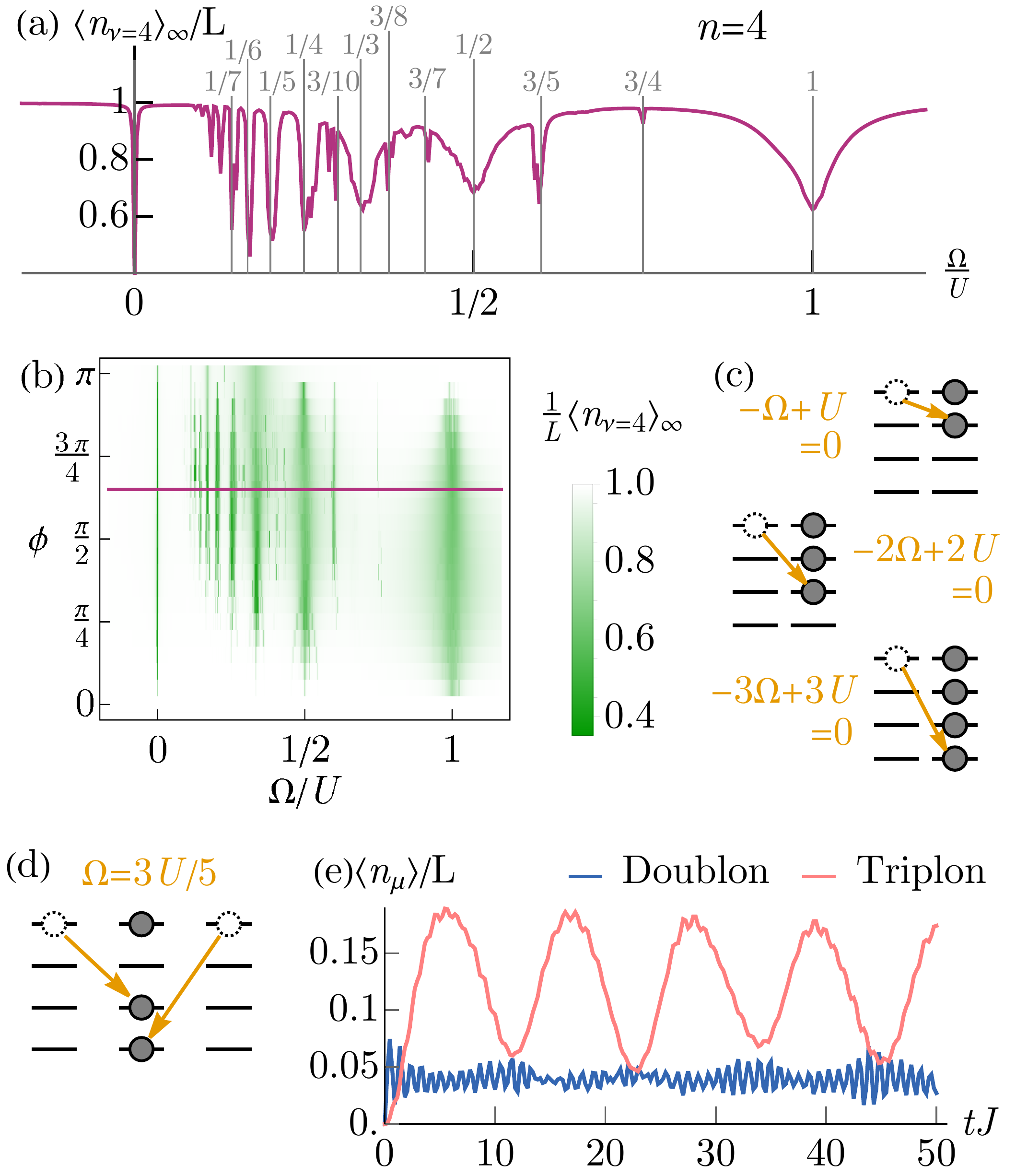}
\centering
\caption{(a) Long-time average population of the initially populated dressed state $\langle \hat{n}_{\nu=4}\rangle_{\infty}$ for an $n=4$ system as a function of $\Omega/U$. The interactions are fixed at $U/J = 30$, the system size at $L=5$ and the flux at $\phi = 2.0$ [we use this flux rather than a simpler value like $\pi/2$ to avoid accidental cancellations of dressed tunneling matrix elements in Eq.~\eqref{eq_DressedTunnelingRate}]. Resonances at $\Omega = \frac{p}{q} U$ are labelled by their $\frac{p}{q}$ values. (b) Long-time average population for the same system across the parameter space of $\Omega/U$ and $\phi$. The purple line corresponds to panel (a). (c) Allowed spin-flip tunneling processes at the $\Omega = U$ resonance. Doublons, triplons and quadruplons can all be formed. (d) Two-step resonant process at the $\Omega = 3U/5$ resonance, forming a triplon. (e) Time-evolution of fraction of lattice sites with exclusively two atoms (Doublon) and three atoms (Triplon) at the $\Omega = 3U/5$ resonance, for parameters $L=6$, $U/J = 30$, $\Omega/J = 18$.}
\label{fig_SUnPopulation}
\end{figure}

Like $n=2$, the initial state used in our study corresponds to all atoms in the highest-energy dressed state $\nu = n$. When near resonances the population of this state will be depleted by interaction-enabled tunneling transferring atoms into other dressed states. To quantify this depletion we numerically compute the long-time average of the population $\langle \hat{n}_{\nu=n}\rangle = \sum_{j}\langle \hat{n}_{j,\nu=n}\rangle$, still in the strongly interacting regime $U/J \gg 1$, across the parameter range of $\Omega / U$ and $\phi$.

Figs.~\ref{fig_SUnPopulation}(a-b) plot the resulting population for an $n=4$ system and fixed $U/J = 30$. There is a variety of multi-body resonances that occur at rational points $\Omega = \frac{p}{q} U$ for $p,q \in \mathbbm{Z}$ and $p \leq q$. For larger $n$ the number and strength of resonances that can be observed on the timescales used in our numerics increases. Processes that were higher-order for $n=2$ can now become direct. For example, the $\Omega = U/2$ resonance is no longer perturbative as an atom can go from $\nu = 4$ to $\nu=2$ via the $J_{4,2}$ tunneling matrix element directly, losing energy $2\Omega$ and compensating it by the interaction energy $U$. There are also resonances for which multiple types of first-order processes occur at once. The $\Omega = U$ resonance can still form a doublon with a single tunneling event, but now can also turn a doublon into a triplon, or a triplon into a quadruplon as shown in Fig.~\ref{fig_SUnPopulation}(c). This yields an interesting behavior where the occupancy of a lattice site does not inhibit the flow of population through it unless all $n$ levels have already been filled, almost analogous to a non-interacting system. It is for this reason that we find good qualitative agreement with the non-interacting case for the currents in Fig.~\ref{fig_SUnChirality}(d) near this resonance. 

Resonances that exclusively create higher-order objects can also be seen. For example the $\Omega = \frac{3}{5} U$ resonance corresponds to a second-order process that starts with three $\nu = 4$ atoms on three adjacent sites, and moves the outer two into the middle site's $\nu = 1, 2$ states with a total cost of $-3\Omega-2\Omega + 3U = 0$, forming a triplon (interaction cost $3U$) as depicted in Fig.~\ref{fig_SUnPopulation}(d). We demonstrate this property in Fig.~\ref{fig_SUnPopulation}(e), which compares the fraction of lattice sites with exclusively two and exclusively three atoms; the latter is much higher at the this resonance. This is why we only see fractions with numerator 3 and not 2 in Fig.~\ref{fig_SUnPopulation}(a); all such fractions correspond to triplon resonances. While there can be resonances of the form $\Omega = \frac{2}{q} U$, these amount to the formation of doublons in different parts of the lattice and are harder to resolve in the numerics for this initial state. All such features can be intuitively understood by writing the system in the diagonal frame and determining which spin-flip tunneling processes conserve energy.

The width of the resonances can be understood from the same arguments as before. All direct resonances will have their widths set by the tunneling matrix elements (for $n=4$ these are $\Omega = U, U/2, U/3$ with rates $J_{4,3}$, $J_{4,2}$, $J_{4,1}$. Higher-order resonances have effective rates that can be estimated with perturbation theory. For example, the $\Omega = U/4$ resonance for $n=4$ requires a two-step process that involves two atoms going down a total of four levels and forming one doublon. This can be done by one atom going from $\nu=4 \to 1$ and another from $4 \to 3$, or both going from $4 \to 2$. The corresponding processes would have rates of $\frac{J_{4,1}}{-3\Omega + U} J_{4,3} \times 4$ and $\frac{J_{4,2}}{-2\Omega+U}J_{4,2} \times 2$ respectively (each having combinatorial factors for the ordering of the steps). The overall rate can be estimated by summing these together. These bare widths are then broadened by an effective bandwidth proportional to the spin-conserving tunneling rates $J_{\nu,\nu}$ of the dressed states involved in the resonance.

\begin{figure}
\includegraphics[width=.48\textwidth]{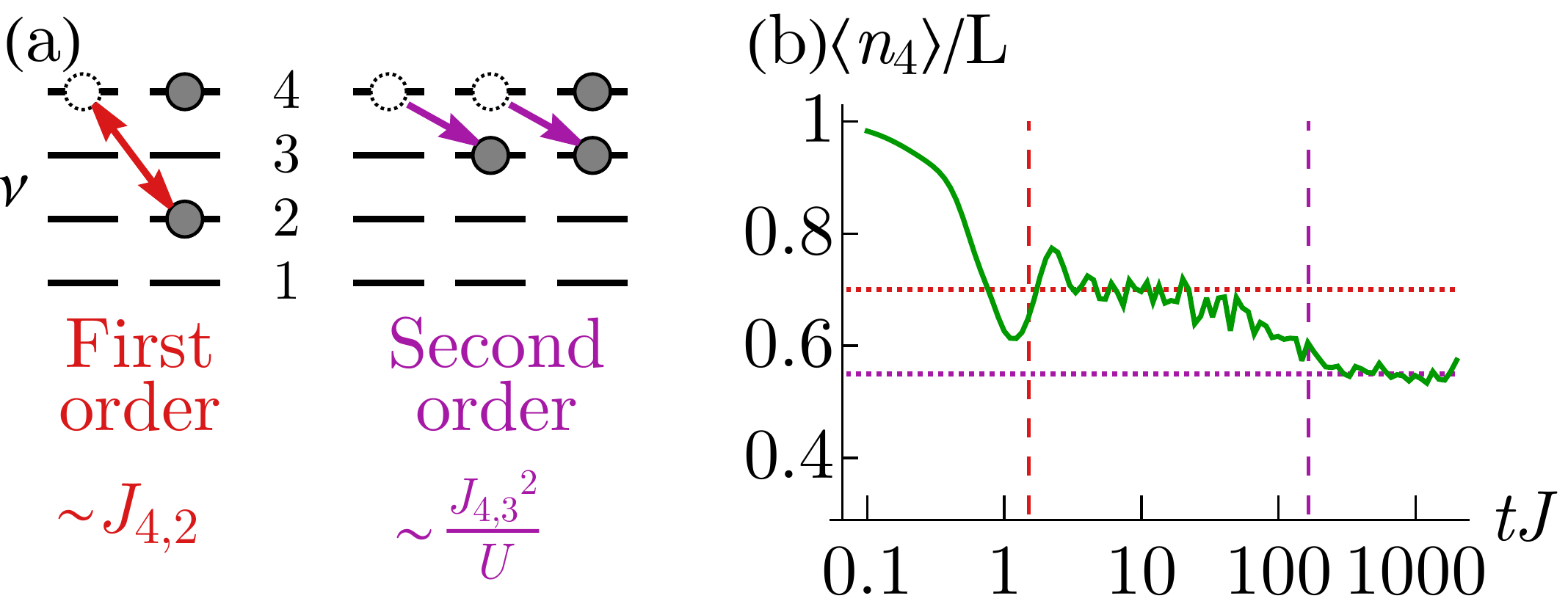}
\centering
\caption{(a) Dominant resonant processes for an $n=4$ system at the $\Omega = U/2$ resonance for our initial state with all atoms in the highest-energy dressed state $\nu = 4$. The typical rates for each process are shown. Other processes with similar rates can also occur, although they require additional moves to become allowed starting from this initial state. (b) Time-evolution of the dressed state population. System size is $L=7$, parameters are $U/J = 30$, $\Omega/J = 15$, $\phi = 2.0$. The system first saturates to an average of $\approx 0.7$ (dotted red line to guide the eye, not a theory prediction), which is close to the prediction of $\frac{2}{3}$ from the first-order resonant process. At later times there is a further saturation to a different average of $\approx 0.54$ (dotted purple line to guide the eye). The associated timescales $t|J_{4,2}|=1$ and $t|J_{4,3}|^2/U=1$ are shown by the dashed red and purple lines respectively.}
\label{fig_SUnThermalization}
\end{figure}

Another interesting insight is that certain resonances can now have multiple allowed processes of different orders happening simultaneously. To demonstrate this behavior we study the time-evolution of the population at a resonance $\Omega = U/2$ for $n=4$. There is a first-order resonance as depicted in Fig.~\ref{fig_SUnThermalization}(a) with interaction-enabled tunneling between $\nu = 4, 2$, analogous to the $\Omega = U$ resonance forming leap-frogging doublons for $n=2$. This resonance dominates the dynamics at short times, leading to a depletion and saturation of the $\nu = 4$ state population on a timescale $t |J_{4,2}| \gtrsim 1$. There is also a second-order resonance that brings two atoms down from $\nu=4 \to 3$, analogous to the $\Omega = U/2$ resonance for $n=2$. This second resonance causes a further depletion in the average population on longer timescales of $t \frac{|J_{4,3}|^2}{U}\gtrsim 1$.

In Fig.~\ref{fig_SUnThermalization}(b) we plot the population's time evolution across a wide temporal range. Two plateaus for the two processes described above are observed to be established on the predicted timescales. The first plateau saturates to a value of $\approx 0.7$, close to the thermal prediction of $\frac{2}{3}$ for the $n=2$ system because the process is identical in nature. The second plateau saturates to a lower value of $\approx 0.54$. This latter average is not yet converged in terms of finite size $L$ (see Appendix~\ref{app_Scaling}) and may require the inclusion of other processes that kick in once the allowed ones in Fig.~\ref{fig_SUnThermalization}(a) have occurred. For example, the $\nu=3$ singlons formed from a second-order resonant process can subsequently undergo first-order resonant tunneling on their own. We expect that for $L \to \infty$ the system settles to an infinite temperature state described by a microcanonical average over all energetically accessible states via both first- and second-order resonant processes.

We note that much of the above discussion does not require a drive with equally spaced eigenenergies $E_{\nu}$. Each distinct energy difference $\Delta E_{\nu,\nu'} = E_{\nu} - E_{\nu'}$ will have an associated tunneling process (and hence first-order resonance) provided that the flux $\phi$ does not cause the associated matrix element $J_{\nu,\nu'}$ to vanish. Higher-order processes can be constructed by composing multiple energetically-allowed moves. Having unequally spaced energies can even help with avoiding different processes acting simultaneously, if one wishes to isolate a specific type of correlated motion.

%%%%%
\section{Experimental implementation}
\label{sec_Implementation}
%%%%%
%%%
\subsection{Drive implementation}
%%%

In this section we will describe how the models we study can be implemented in optical lattice experiments using ultracold alkaline earth atoms. Such systems feature long-lived ground and metastable excited states with zero electronic angular momentum, rendering them robust to perturbations such as stray magnetic fields. Experiments with alkaline earth atoms loaded in 3D  optical lattices have also shown coherence times longer than $10$s~\cite{Hutson2019}, which enables the resolution of long-time average behavior.

For the simplest $n=2$ case, the spin states $m \in e, g$ can be represented by long-lived clock states in a magic wavelength lattice. The laser drive can be realized with a direct optical transition~\cite{fallaniSOCSU2Optical2016,Kolkowitz2017,bromley2018dynamics}. The flux arises from the projection of the clock laser wavevector $\vec{k}_c$ onto the lattice, $\phi = a |\vec{k}_{c}| \cos(\theta)$ with $a$ the lattice spacing and $\theta$ the angle of the laser to the tunneling direction. We consider lattice depths of $\sim 10 E_r$ along the tunneling direction and $\sim 100 E_r$ along transverse directions (with $E_r$ the recoil energy), leading to interactions on the order of $U \sim 1$ kHz and tunneling on the order of $J \sim 10-100$ Hz for typical candidate AEAs such as $^{87}$Sr. The laser drive Rabi frequency would need to be comparable to the interactions, $\Omega \sim 1$ kHz.

For generic $n \geq 2$, the spin states $m \in \{ -S \dots S\}$ can also be realized as the levels of a hyperfine manifold, for which the drive can be implemented using Raman lasers~\cite{fallaniChiralCurrents2015,hanToblerone2019,song2016spin,song2018observation,song2019observation}. The lattice depth, tunneling and interaction strengths should be on the same scales as listed above; for the drive, the two-photon Rabi frequency of the Raman couplings would need to be on the order of $\sim 1$ kHz to match the interactions. The specific form of the couplings depends on the beam configuration. We will first discuss a simple scheme realizing ``nearest-neighbour'' couplings $\Omega_{m,m+1}$ along the synthetic dimension. Afterwards, we will show a more complex scheme generating the $\hat{S}^{x}$ drive with equally-spaced eigenenergies that we studied in preceding sections.

\begin{figure*}
\includegraphics[width=1\textwidth]{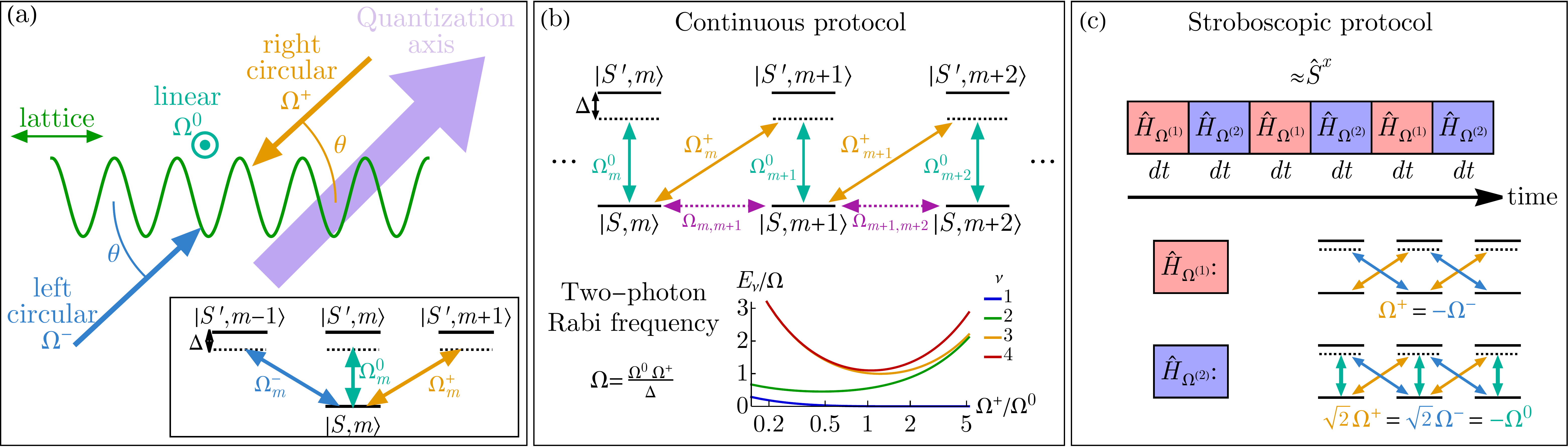}
\centering
\caption{(a) General schematic for realizing the SU($n$) system using a Raman coupling in a lattice with alkaline-earth atoms. Far-detuned Raman beams with different polarizations illuminate the lattice. The one-photon couplings $\Omega_{m}^{0}$, $\Omega_{m}^{\pm}$ are all $m$-dependent due to Clebsch-Gordan coefficients, but proportional to overall magnitudes $\Omega^{0}$, $\Omega^{\pm}$. The linearly-polarized $\Omega^{0}$ beam propagates perpendicular to the lattice direction, while the right- and left-circular beams $\Omega^{+}$, $\Omega^{-}$ point in equal and opposite directions at an angle $\theta$ to the lattice direction. (b) Protocol for generating a ``nearest-neighbour'' drive coupling along the synthetic dimension. The linear and right-circular beams are used to realize effective two-photon couplings $\Omega_{m,m+1}$. The overall intensity of the couplings is set by $\Omega = \Omega^{0} \Omega^{+} / \Delta$. The bottom plot shows the energies of the on-site drive eigenstates as a function of $\Omega^{+}/\Omega^{0}$, demonstrating that the dressed state spectrum can be tuned. (c) Protocol for generating a spin-$S$ $\hat{S}^{x}$ type coupling. Two different laser configurations $\hat{H}_{\Omega^{(1)}}$ and $\hat{H}_{\Omega^{(2)}}$ are stroboscopically alternated. They realize effective drives of $(\hat{S}^{x})^2$ and $-(\hat{S}^{x})^2 + \hat{S}^{x}$ respectively, yielding the desired $\hat{S}^{x}$ provided the stroboscopic rate is faster than the tunneling and interactions.}
\label{fig_DriveImplementation}
\end{figure*}

We show a sample configuration of Raman beams in Fig.~\ref{fig_DriveImplementation}(a). Our proposed drive schemes will use some/all of these beams. There is a linear polarized beam (labelled $\Omega^{0}$) orthogonal to the lattice direction, and a pair of right-circular ($\Omega^{+}$) and left-circular ($\Omega^{-}$) polarized beams pointed in equal and opposite directions at an angle $\theta$ to the lattice direction, along the quantization axis of the system. The beams couple a ground electronic hyperfine manifold of $n$ states with total angular momentum $S=(n-1)/2$ and projection $m \in \{-S, \dots, S\}$, which will be used as the synthetic dimension, to an excited electronic hyperfine manifold with total angular momentum $S'$. Each beam has wavevector magnitude $|\vec{k}_L|$ and a detuning $\Delta$. The on-site single-particle Hamiltonian describing this laser configuration is,
\begin{equation}
    \hat{H}_{\mathrm{Raman}} = \hat{H}_{\Delta} + \hat{H}_{\Omega^{0}} + \hat{H}_{\Omega^{+}} + \hat{H}_{\Omega^{-}},
\end{equation}
where $\hat{H}_{\Delta}$ is the detuning,
\begin{equation}
    \hat{H}_{\Delta} = \Delta \sum_{m} \ket{S',m}\bra{S',m},
\end{equation}
while the other terms are the laser couplings,
\begin{equation}
\begin{aligned}
    \hat{H}_{\Omega^{0}} &= \sum_{m} \bigg(\Omega^{0}_m \ket{S,m} \bra{S',m} + h.c.\bigg),\\
    \hat{H}_{\Omega^{+}} &= \sum_{m} \bigg(\Omega_{m}^{+}\ket{S,m}\bra{S',m+1} + h.c.\bigg),\\
    \hat{H}_{\Omega^{-}} &=\sum_{m} \bigg(\Omega^{-}_{m} \ket{S,m} \bra{S',m-1} + h.c.\bigg).\\
\end{aligned}
\end{equation}
The single-photon coupling matrix elements are
\begin{equation}
\begin{aligned}
    \Omega^{0}_{m} &= \Omega^{0} \langle S,m;1,0|S',m\rangle,\\
    \Omega^{\pm}_{m} &=\Omega^{\pm} \langle S,m;1,\pm1 |S',m\pm 1 \rangle e^{\pm ij\phi},
\end{aligned}
\end{equation}
where $\Omega^0$, $\Omega^{\pm}$ are overall strengths set by the laser power and $\langle S,m;1,0|S',m \rangle$, $\langle S,m;1,\pm 1 |S',m\pm 1 \rangle$ are Clebsch-Gordan coefficients. The right- and left-circular polarized lasers also have spin-orbit coupling phases $e^{+i j \phi}$ and $e^{-i j \phi}$ respectively. The flux is $\phi  = a |\vec{k}_{L}| \cos\theta$. In principle each beam can have an independent angle to the lattice direction (and hence spin-orbit coupling phase); we choose the configuration in Fig.~\ref{fig_DriveImplementation}(a) because it is simpler while still allowing implementation of the desired schemes.

We show how to generate simple nearest-neighbour couplings along the synthetic dimension in Fig.~\ref{fig_DriveImplementation}(b). This scheme uses the linear and right-circular polarized beams $\hat{H}_{\Omega^{0}}$, $\hat{H}_{\Omega^{+}}$. For large detuning $|\Delta| \gg |\Omega^0_{m}|, |\Omega^+_{m}|$ we adiabatically eliminate the excited manifold $S'$ and obtain an effective two-photon coupling between the ground hyperfine states given by $\Omega_{m,m+1} = \Omega^0_{m+1} \Omega^{+}_{m}/\Delta$ (in the gauged frame). Each state also experiences Stark shifts given by $\Omega_{m,m} =(|\Omega^0_{m}|^2+|\Omega^{+}_{m}|^2)/\Delta$. While these couplings are not the same as the ones we studied in the previous sections, similar resonant physics can still be obtained because the main factor determining the resonances is the eigenspectrum $E_{\nu}$ of the drive. The overall energy scale of $E_{\nu}$ can be defined as $\Omega = \Omega^{0} \Omega^{+}/\Delta$. The splitting between the drive eigenstates can be tuned by controlling the ratio of the overall laser intensities $\Omega^+/\Omega^0$. Fig.~\ref{fig_DriveImplementation}(b) shows the spectrum for a sample system with $n = 4$ levels in both the ground and excited manifolds ($S = S'= 3/2$). By tuning the intensity ratio, one can change the dressed state energies and thus the permitted resonances.

We can also realize the $\hat{S}^{x}$ drive which has equally-spaced eigenenergies and couplings $\Omega_{m,m+1}=\Omega_{m+1,m}=\frac{\Omega}{2} \sqrt{(S-m)(S+m+1)}$, with a stroboscopic protocol as discussed in Ref.~\cite{perlinSUn2021}. We depict this scheme in Fig.~\ref{fig_DriveImplementation}(c). The scheme employs two beam configurations which are alternated for equal time intervals $dt$ much shorter than the timescales of the lattice dynamics. Specifically, we use $\hat{H}_{\Omega^{(1)}}$ and $\hat{H}_{\Omega^{(2)}}$ (still with a large detuning throughout), given by:
\begin{equation}
\begin{aligned}
    \hat{H}_{\Omega^{(1)}} &= \hat{H}_{\mathrm{Raman}} \text{  with } \>\>\>\>\>\Omega^{0} = 0, \>\> \Omega^{+} = - \Omega^{-},\\
    \hat{H}_{\Omega^{(2)}} &= \hat{H}_{\mathrm{Raman}} \text{  with } \>\>\>\>\>\Omega^{+} = \Omega^{-} = - \sqrt{2} \Omega^{0}.
\end{aligned}
\end{equation}
These effectively implement (after adiabatically eliminating the excited manifold) the following drive couplings:
\begin{equation}
\begin{aligned}
    \hat{H}_{\Omega^{(1)}} &\approx +\Omega(\hat{S}^{x})^2,\\
    \hat{H}_{\Omega^{(2)}} &\approx -\Omega(\hat{S}^{x})^2 + \Omega\hat{S}^{x},\\
    \Omega &= \frac{8(\Omega^{+})^2}{15 \Delta}.
\end{aligned}
\end{equation}
Here $\Omega$ is the effective two-photon Rabi frequency for this implementation, and the flux is still set by $\phi = a |\vec{k}_{L}|\cos \theta$. Stroboscopically alternating these two Hamiltonians yields an effective drive $\Omega \hat{S}^{x}$ as desired.

%%%
\subsection{State preparation and measurement}
%%%

\begin{figure*}
\includegraphics[width=0.8\textwidth]{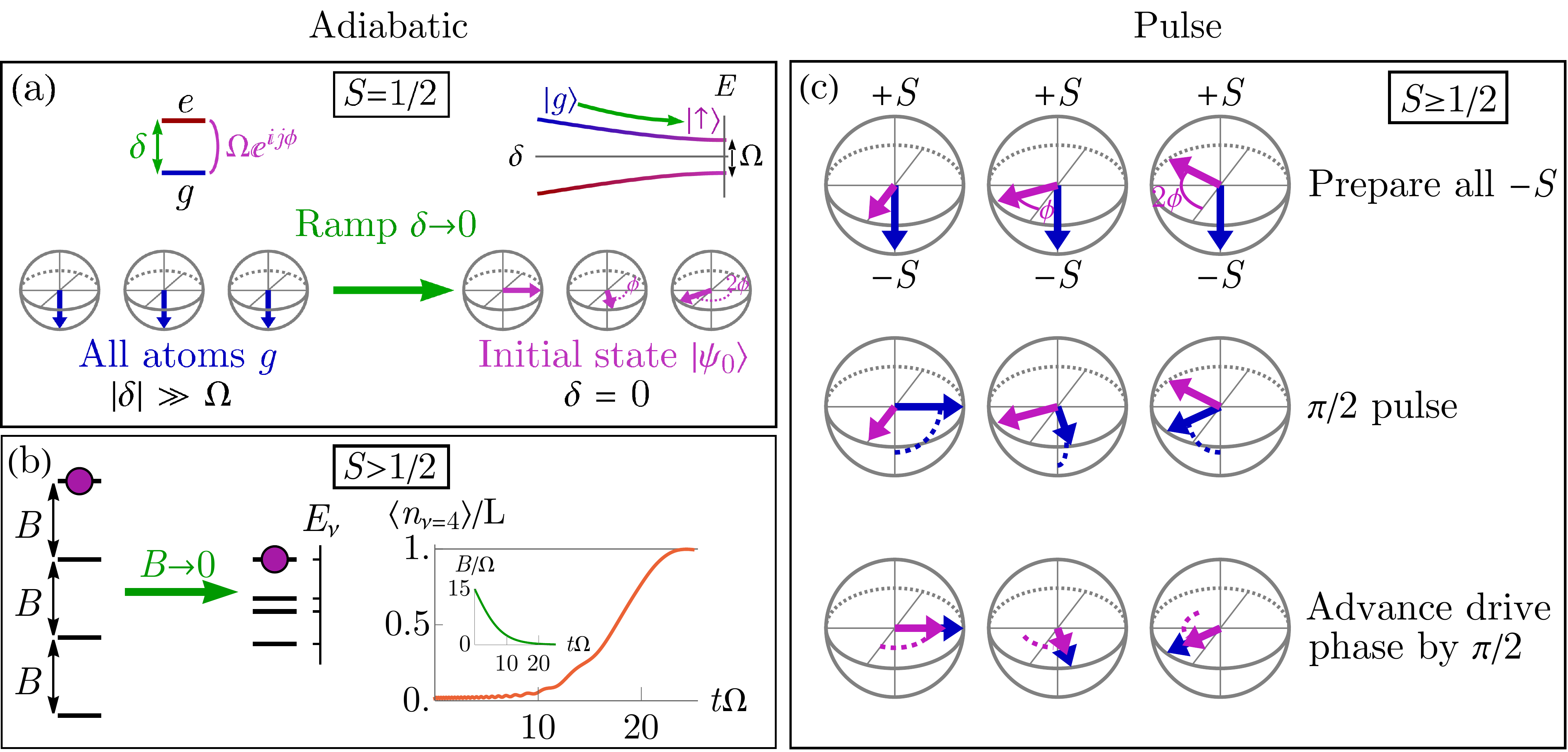}
\centering
\caption{(a) State preparation using a single-particle adiabatic ramp for $n=2$ (spin $S = 1/2$). Atoms are initialized in the lab frame spin state $g$, and a laser coupling is turned on with a large negative detuning $-\delta \gg \Omega$. Ramping $\delta \to 0$ sweeps the system into the desired initial state $\ket{\psi_0}$ with every lattice site in the highest-energy eigenstate $\uparrow$ of the drive. (b) Similar adiabatic preparation for an $n>2$ system ($S > 1/2$). The detuning is replaced by e.g. a magnetic field splitting of hyperfine states proportional to a field strength $B$, which is ramped to zero. The plot shows a numerical simulation of such a ramp for an $n = 4$ system, using the coupling scheme in Fig.~\ref{fig_DriveImplementation}(b) for sample parameters of $\Omega^{+}/\Omega^{0} = 3$, $\Omega / J = 2$ and $U/J = 30$. The inset shows the ramp profile. The highest-energy drive eigenstate $\nu = 4$ is prepared with $> 99 \%$ fidelity. (c) State preparation protocol using a pulse and phase skip sequence, which works for a system with an $\hat{S}^{x}$ drive and any $n \geq 2$. Blue arrows denote the spin direction of each atom, while purple arrows denote the laser drive direction. Atoms are first prepared in a single lab frame spin state $m = -S$ without any detuning or field splitting. The laser drive is turned on and used to implement a $\pi/2$ pulse, rotating the spins onto the equator of the Bloch sphere (or the generalized Bloch sphere of the fully symmetric manifold for $n > 2$). The drive phase is then skipped ahead by $\pi/2$ with e.g. a fast detuning pulse, which advances its axis of rotation to be aligned with the spin state. This prepares the desired $\ket{\psi_0}$.}
\label{fig_StatePreparation}
\end{figure*}

Preparing the initial state $\ket{\psi_0}$ is possible with either a single-particle adiabatic ramp, or a pulse sequence. The ramp protocol is shown for the simplest $n = 2$ case in Fig.~\ref{fig_StatePreparation}(a). The system is initialized with all atoms in the lab frame spin state $g$. Then, the drive laser is turned on, with an initially large negative detuning $\delta < 0$ corresponding to a Hamiltonian term $\hat{H}_{\delta} = \frac{\delta}{2}\sum_{j}\left(\hat{n}_{j,e} - \hat{n}_{j,g}\right)$, which is subsequently ramped to $\delta = 0$. This adiabatically transfers every atom to the dressed state $\uparrow$, realizing the required spiral state (in the lab frame). To prevent tunneling dynamics during state preparation, one can start with a very deep lattice such that $J$ is negligible on the timescale of the ramp, and reduce the depth after preparation is done.

For $n > 2$, state preparation is still possible with a single-particle adiabatic ramp. To do so all of the $n$ hyperfine levels must be split in energy, which is not possible with a single laser detuning as was shown for $n=2$, but can be realized with for example a magnetic field ramp,
\begin{equation}
    \hat{H}_{B}(t) = B (t) \sum_{j,m}m\> \hat{n}_{j,m}.
\end{equation}
The scheme starts with a spin polarized state with all atoms in a single hyperfine level $m=+S$ [the highest-energy state provided $B(0) > 0$] and a large field $|B(0)| \gg |\Omega|$. The field is then ramped down to $B = 0$, yielding an eigenstate of the drive with the highest energy $\nu = n$. To test this numerically, we use the sample coupling scheme in Fig.~\ref{fig_DriveImplementation}(b). We fix the amplitudes $\Omega^0$, $\Omega^+$ and numerically simulate an adiabatic ramp of $\hat{H} + \hat{H}_{B}(t)$ for a candidate ramp profile in Fig.~\ref{fig_StatePreparation}(b), showing that one can initialize most atoms in a single drive eigenstate as required.

As an alternative to an adiabatic ramp, for an $\hat{S}^{x}$ drive [realized using a direct laser coupling for $n=2$ or the stroboscopic scheme in Fig.~\ref{fig_DriveImplementation}(c) for $n>2$] one can prepare the initial state $\ket{\psi_0}$ with a single laser pulse followed by a phase skip. This preparation is shown in Fig.~\ref{fig_StatePreparation}(c) in the lab frame. One initializes all atoms in a single $m = -S$ state, without any detuning or field splitting. The laser drive is then turned on for a time $t \Omega = \frac{\pi}{2}$, realizing a $\pi/2$ pulse that rotates each spin onto the equator of the spin-$S$ Bloch sphere. The phase of the laser drive is finally skipped ahead by $\pi$/2, which aligns its axis of rotation with the spin. This realizes the desired state $\ket{\psi_0}$ with each atom in the highest-energy drive eigenstate.

For either preparation scheme, measurements of the initially populated dressed state can be performed by reversing the state preparation and measuring the population of the initially loaded hyperfine state using standard spectroscopy protocols. Measuring the currents is possible using spin-resolved time-of-flight measurements~\cite{fallaniChiralCurrents2015}, exploiting that the current of a given leg $m$ is given by $\sum_{k} \sin(k a) \hat{n}_{k,m}$.

%%%
\subsection{Imperfections}
%%%

Here we provide a discussion of possible experimental imperfections, considering non-unit-filling, external trapping potentials, imperfect alignment of the lasers, and unequal tunneling rates.

\begin{figure}
\includegraphics[width=0.5\textwidth]{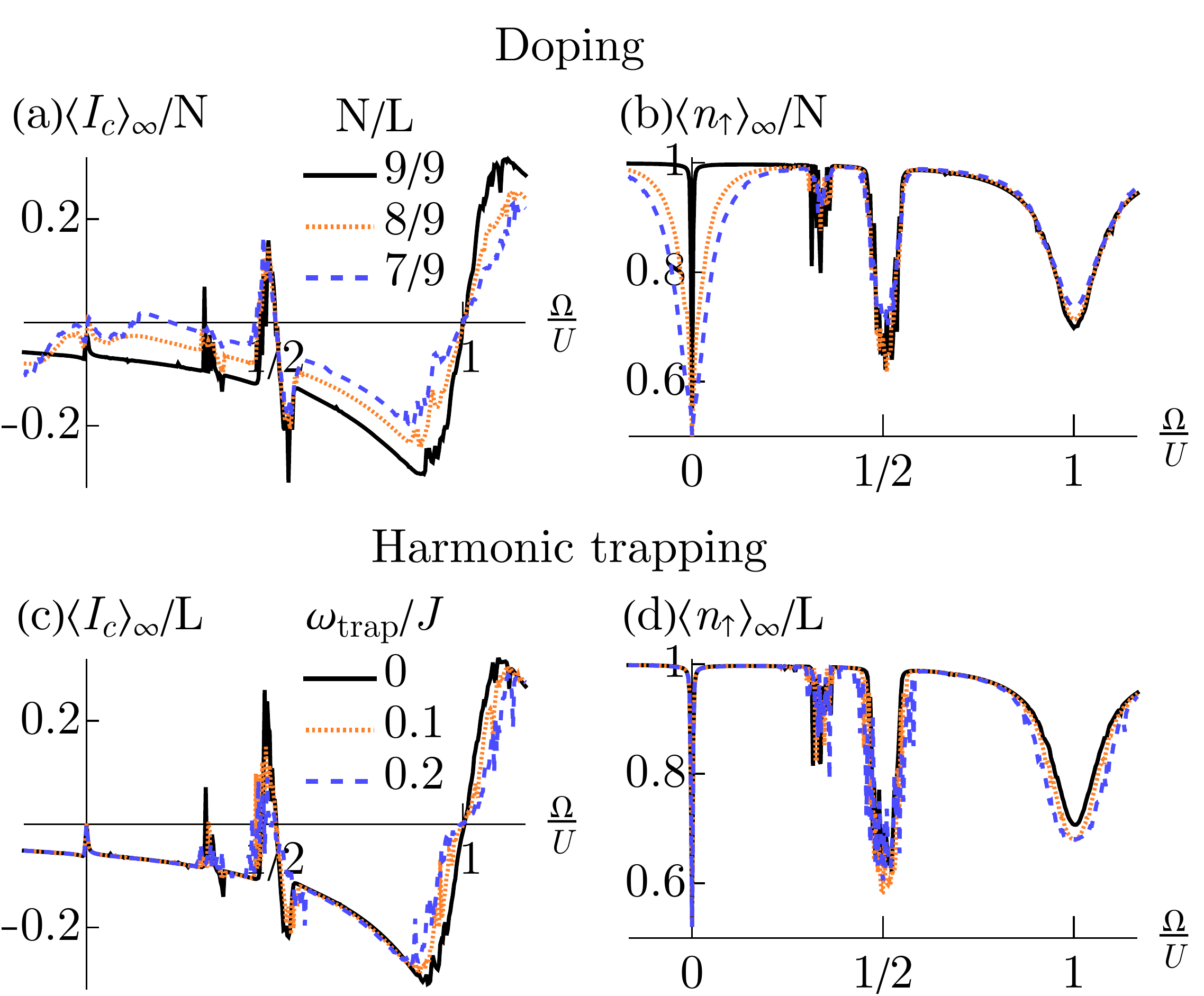}
\centering
\caption{
(a) Average chiral current for the $n=2$ system with holes in the initial state. The parameters are $L=9$, $U/J = 30$, $\phi = \pi/2$. Time-averaging is done out to $tJ = 500$. The $N/L=8/9$ curves have a single hole initially at lattice site $j=1$, while the $N/L = 7/9$ curves have holes initially at sites $j=1,3$. (b) Average dressed state population for the same setup.
(c) Average chiral current for the $n=2$ system in the presence of a harmonic trap $\hat{H}_{\mathrm{trap}}$. We use the same parameters as (a), except with open boundaries and no holes. (d) Average dressed state population for the same setup.}
\label{fig_Imperfections}
\end{figure}

One potential source of error is an imperfect filling fraction $N/L < 1$ at finite temperatures. However, the presence of holes in the initial state is not a major concern for the dynamical signals we study. Holes will allow for more tunneling (thus broadening the resonances), but will not inhibit the interaction-enabled processes completely. We verify this in Fig.~\ref{fig_Imperfections}(a-b), which plots the average chiral current and dressed state population for $n=2$ in the presence of one or more holes in the initial state (assuming the atoms that are present still get prepared in the highest-energy dressed state). The higher-order resonant features become weaker and broader while the $\Omega = 0$ resonance (corresponding to the non-interacting resonance) becomes stronger; however, all of the features discussed previously can still be resolved.

Another physical mechanism which affects the resonances is harmonic trapping due to finite laser beam waist, which can be captured by a simple quadratic spin-independent potential:
\begin{equation}
    \hat{H}_{\mathrm{trap}} = \omega_{\mathrm{trap}} \sum_{j} (j-j_0)^2 \sum_{m}\hat{n}_{j,m},
\end{equation}
with $\omega_{\mathrm{trap}}$ the trap energy and $j_0 = (L+1)/2$ the center of the lattice. A harmonic trap will also cause broadening since the energy cost for resonant processes will become site dependent, but should not destroy the predicted resonances provided that the site-to-site trap energy differences near the center of the lattice are weak compared to the tunneling rates. This is confirmed in Fig.~\ref{fig_Imperfections}(c-d) which show the chiral current and dressed state population for different trap energies.

Yet another potential error source is imperfect control over the laser alignment angles. However, such misalignment will only cause small discrepancies in the flux $\phi$, which should not significantly change the long-time dynamical behavior so long as the dressed state tunneling rates $J_{\nu,\nu'}$ relevant to a desired resonance do not completely vanish if $\phi$ is perturbed.

Finally, we have assumed equal tunneling rates $J$ for all of the bare spin states $m$. Any deviation from this regime due to e.g. non-magic lattices does not significantly alter our results even if the shifts are non-negligible. The dressed state tunneling rates $J_{\nu,\nu'}$ are already sums over the independent tunneling matrix elements of the bare spin states in Eq.~\eqref{eq_DressedTunnelingRate}. An $m$-dependent tunneling rate would enter into the sum and modify the dressed state matrix elements $J_{\nu,\nu'}$, but will not qualitatively alter the physics that we describe provided that the relevant $J_{\nu,\nu'}$ do not vanish.

%%%%%
\section{Conclusions and outlook}
%%%%%

We investigated the quench dynamics of driven and strongly interacting SU$(n)$ fermions with one atom per site in synthetic flux ladders. We showed the emergence of rich interaction-induced multi-body resonances at specific values of the Rabi drive, which induce particle flow and generate chiral currents with non-trivial patterns. Ladders involving larger number of internal levels exhibit more prolific spectra of resonances that manifest on shorter timescales (set by the tunneling rate) because of the presence of more types of energetically favourable processes. The resonances exhibit kinetically constrained quantum dynamics determined by the allowed types of tunneling, which can be intuitively understood in the diagonal frame. One advantage in this setup is that the atoms are highly mobile while still being subject to non-trivial density-dependent effects, which can allow the natural generation of exotic correlated states through dynamical evolution without needing to reach ultra-cold temperatures to prepare ground states or to engineer complicated many-body Hamiltonians or initial conditions.

An interesting future direction is to compare our theory results to systems that emulate more conventional 2D materials. Such systems would have a synthetic dimension coupling that is homogeneous rather than $m$-dependent (i.e. treating the synthetic dimension as a proper lattice dimension on its own right). Furthermore, unlike the all-to-all interactions in the synthetic direction due to SU(n) symmetry in our system, the interactions in real materials will be short-ranged in both directions. Theoretical comparisons to such systems are relevant to studies of the fractional quantum Hall effect. It is likely that the fractional resonances we see will still survive with shorter range interactions because we work in a basis of dressed states determined by diagonalizing the couplings along the synthetic dimension. These dressed states are delocalized along that dimension, which will cause even short-range interactions to effectively become long-range when the interaction Hamiltonian is written in the dressed basis.

The ability to isolate special types of resonant processes with this system also opens pathways to several other general applications in quantum simulation. For example the ability to exclusively form triplons without creating doublons enables the clean study of higher-order collisional interactions. By bringing the system to the corresponding resonance one can measure the effect of coherent three-body interactions~\cite{goban2018emergence} (which would shift the position of the resonance), or the associated incoherent loss rates (which would lead to faster atom number decay at the resonance). Such correlated processes are also integral to the simulation of dynamical gauge theories. One can understand the restriction of only forming triplons as an effective symmetry constraint on the permitted site occupancy. As a further extension, if the drive spectrum $E_{\nu}$ exhibits degeneracies one can have resonantly coupled subspaces with more than one state per site. This can lead to effective non-Abelian extensions of density-dependent tunneling models, where the mobile particles themselves have effective spin degrees of freedom within their dynamically constrained Hilbert space. Generating such a system is possible with a drive that resembles a $(\hat{S}^{x})^2$ operator for example, which can be realized using the beam configurations discussed in the previous section following Ref.~\cite{perlinSUn2021}. However, we delegate the study of such systems to future work.

Another avenue for future work is the study of these systems in higher spatial dimensions. We considered 1D chains (hence 2D ladders when including the synthetic dimension) for numerical simplicity. However, it is straightforward to extend these types of schemes to 2D or even 3D lattice configurations using 3D optical lattice experiments, simply by reducing lattice depth along more than one dimension. Conventional Raman coupling schemes translate in a straightforward manner; depending on the angle of the beams, each lattice direction will have its own associated flux $\phi_x$, $\phi_y$, etc.. All of the methodology that we have used will remain applicable, with separate spin-flip tunneling matrix elements along each dimension in the diagonal frame. Moreover, one can even tune the fluxes to allow certain types of resonant processes along one dimension but not another by tuning the associated tunneling matrix elements by the corresponding flux. Finally, all of the physics we have considered has been focused on measuring collective observables of the system (in the appropriate frame). State-of-the-art tools like optical tweezers and quantum gas microscopes for AEAs allow for the preparation and site-resolved measurement of quench dynamics for non-uniform initial states~\cite{urech2022tweezer,young2022tweezer,madjarov2020tweezer,saskin2019tweezer}. A particularly promising aspect is hybrid systems using both tweezers and lattices, which can allow the preparation of specific multi-atom configurations that undergo resonant dynamics (like the leap-frog doublons). This possibility opens the avenue to study their propagation, transport,
collisional properties and relaxation in a spatially resolved fashion.

\textbf{Acknowledgements.}
We thank Aaron W. Young and Tobias Bothwell for a careful reading and comments on the manuscript. This work is supported by the AFOSR grants FA9550-18-1-0319 and FA9550-19-1-0275, by the NSF JILA-PFC PHY-1734006, QLCI-OMA -2016244, by the U.S. Department of Energy, Office of Science, National Quantum Information Science Research Centers Quantum Systems Accelerator, and by NIST.

\bibliographystyle{unsrt}
\bibliography{ResonantBiblio.bib}
\clearpage
\begin{appendix}

\section{Numerical convergence}
\label{app_Scaling}

\begin{figure}[h]
\includegraphics[width=.48\textwidth]{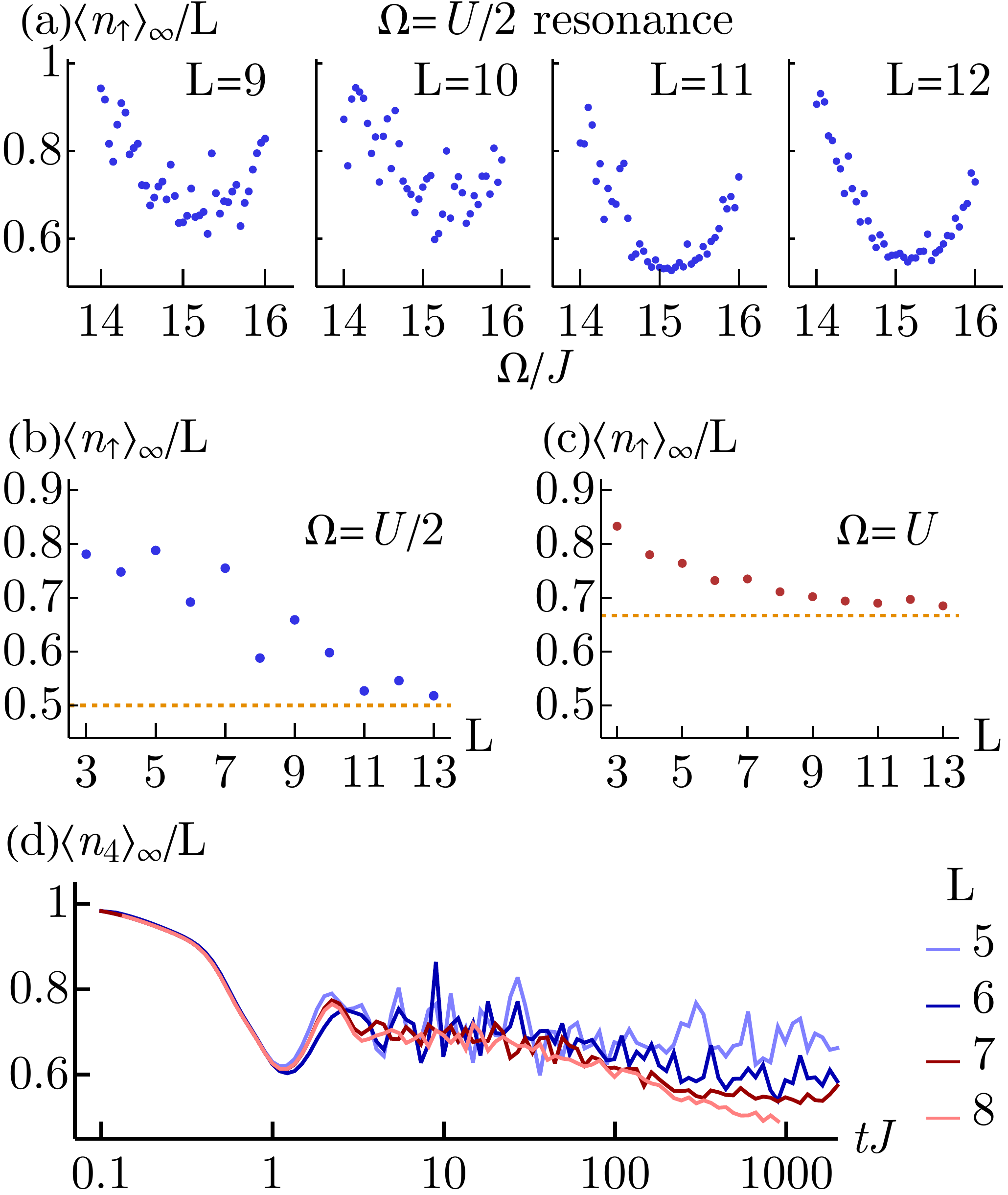}
\centering
\caption{(a) Long-time average population $\langle \hat{n}_{\uparrow}\rangle_{\infty}$ for the $n=2 $system near the $\Omega=U/2$ resonance, for different system sizes $L$. The parameters are $U/J = 30$, $\phi = \pi/2$, time evolution is done to $tJ = 1500$ and averaging is done over the second half of the evolution, $tJ = 750$ to $1500$. (b) Long-time average population at the $\Omega = U/2 $ resonance in terms of system size, fixing $\Omega / J =15.15$ (where the peak appears to be centered). The orange dashed line is the thermal prediction of $1/2$. (c) Same as (b) but for the $\Omega = U$ resonance, fixing $\Omega/J=30.15$. (d) Long-time average population for the $n=4$ system at the $\Omega = U/2$ resonance for different system sizes $L$. The parameters are $U/J = 30$, $\phi = 2.0$.}
\label{fig_Scaling}
\end{figure}

In this appendix we provide numerical scaling with system size $L$ for some of the results presented in the paper. Fig.~\ref{fig_Scaling}(a) shows the long-time average population for the $n=2$ system near the $\Omega = U/2$ resonance, in analogy with the middle panel of Fig.~\ref{fig_SU2Thermalization}(a). The profile of the peak becomes more smooth and the height increases for larger $L$. In Fig.~\ref{fig_Scaling}(b) we plot the scaling of the peak height with $L$, showing that it saturates towards a finite value. This value is in good agreement with the predicted thermal average of $1/2$, as derived in Appendix~\ref{app_Thermal}. Fig.~\ref{fig_Scaling}(c) shows the scaling of the peak height for the $\Omega = U$ resonance instead, for which the agreement with the prediction of $2/3$ is also reasonable.

In Fig.~\ref{fig_Scaling}(d) we show scaling of the transient evolution for the excited state population of the $n=4$ system, same as Fig.~\ref{fig_SUnThermalization}(b). The average of the first plateau in these profiles is in good agreement with the value of $\langle \hat{n}_{4}\rangle_{\infty}/L = 2/3$. The average of the second plateau does not yet look converged for the accessible system sizes, although it also appears to be saturating towards a thermal average. The main text figure does not use the largest system size ($L=7$ instead of $L=8$) so that the two plateaus are clearly visible.

\section{Different initial state}
\label{app_DifferentInitial}

For the $n=2$ system we study dynamics of an initial state with all atoms in the $\uparrow$ dressed state, corresponding to a delocalized superposition of the spin states $e$, $g$ on every site, $\ket{\psi_0} = \prod_{j}\hat{a}_{j,\uparrow}^{\dagger}\ket{0}=\prod_j\frac{1}{\sqrt{2}}(e^{i j \phi/2}\hat{c}_{j,e}^{\dagger}+e^{-i j \phi/2}\hat{c}_{j,g}^{\dagger})\ket{0}$ in the lab frame. The resonances we see are not exclusive to this state. Every initial condition can be written in the diagonal frame as a superposition of Fock states $\{\ket{n_{1,\uparrow},n_{1,\downarrow},n_{2,\uparrow},n_{2,\downarrow},\dots }\}$ with $n_{j,\nu}$ the number of $\nu\in\{\uparrow,\downarrow\}$ dressed atoms on site $j$. Our initial condition is $\ket{\psi_0} = \ket{1,0,1,0,1,0,\dots}$. A different initial condition can also undergo resonant dynamics, provided that at least one of its Fock state components has some allowed moves as shown in Fig.~\ref{fig_Population}(a).

\begin{figure}[h]
\includegraphics[width=.4\textwidth]{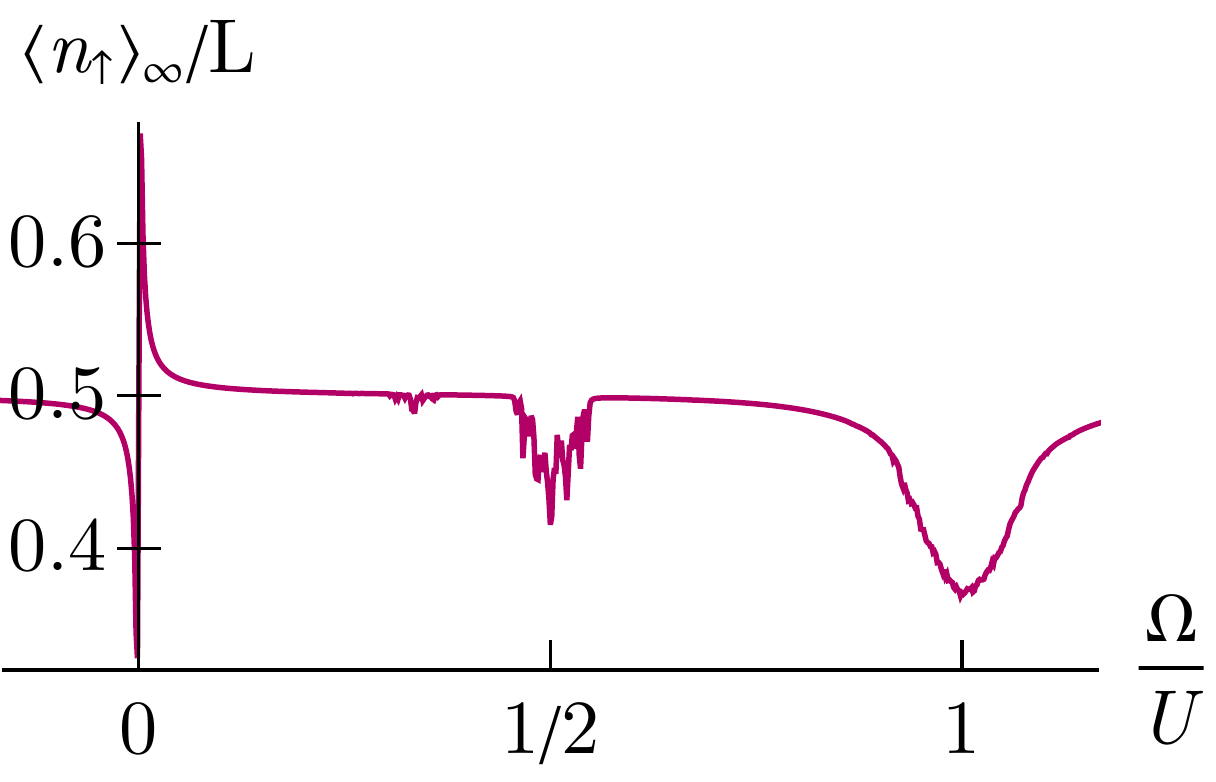}
\centering
\caption{
Long-time average population $\langle \hat{n}_{\uparrow}\rangle_{\infty}$ for the $n=2$ system with the different initial state $\ket{\tilde{\psi}_0}$ from Eq.~\eqref{eq_DiffInitial}. The parameters are $L=9$, $U/J = 30$, $\phi = \pi/2$, and time averaging is done out to $tJ = 500$.
}
\label{fig_DifferentInitial}
\end{figure}

To demonstrate this property, in Fig.~\ref{fig_DifferentInitial} we plot the long-time average population $\langle \hat{n}_{\uparrow}\rangle_{\infty}$ for a different initial state,
\begin{equation}
    \ket{\tilde{\psi}_0} = \prod_{j} \hat{c}_{j,e}^{\dagger}\ket{0},
\end{equation}
corresponding to all atoms starting in the $e$ spin state (thus one leg of the ladder). We see that the resonances are still present, just reduced in magnitude. The reduction happens because only some components of the state can undergo dynamics. For example, if we have $L=2$ this initial state can be written in the diagonal frame Fock basis as,
\begin{equation}
\label{eq_DiffInitial}
\begin{aligned}
    \ket{\tilde{\psi}_0}|_{L=2} &= \prod_{j=1}^{2}\frac{1}{\sqrt{2}}\left(\hat{a}_{j,\uparrow}^{\dagger} + \hat{a}_{j,\downarrow}^{\dagger}\right)\ket{0}\\
    &= \frac{1}{2}\big(\ket{1,0,1,0} + \ket{1,0,0,1}\\
    &\>\>\>\>\>\>\>+\ket{0,1,1,0}+\ket{0,1,0,1}\big).
\end{aligned}
\end{equation}
At the $\Omega = U$ resonance the $\ket{1,0,1,0}$ state can undergo dynamics via resonant tunneling while the rest will remain frozen due to the constraints. Larger systems and higher-order resonances will lead to different fractions of states that are "allowed" to move, but the system will have a response provided that this fraction is non-negligible.

\section{Thermal average}
\label{app_Thermal}

Here we show how the thermal average of the $\langle \hat{n}_{\uparrow}\rangle$ observable for the $n=2$ system can be predicted for resonances of the form $\Omega = \frac{1}{q}U$ with $q \in \{1,2,3,\dots\}$. At resonance, the full Hilbert space of the system has a subspace of $n_{\mathrm{res}}$ Fock states $\{\ket{\phi_n}\}$ which have the same energy as the initial state, $\bra{\phi_n} \hat{H} \ket{\phi_n} = \bra{\psi_0} \hat{H} \ket{\psi_0} = \frac{L \Omega}{2}$. During dynamics the system equilibrates by exploring this subspace. Upon equilibrating, it reaches a thermal state described by a microcanonical ensemble within the subspace. Since this ensemble is effectively at infinite temperature, the equivalent density matrix is given by $\rho_{\mathrm{th}}= \hat{P}_{\mathrm{res}}/n_{\mathrm{res}}$ where $\hat{P}_{\mathrm{res}}=\sum_{n}\ket{\phi_n} \bra{\phi_n}$ projects onto the accessible subspace. The thermal value of the observable is then $\langle\hat{n}_{\uparrow}\rangle_{\infty}= \text{tr}(\hat{P}_{\mathrm{res}}\hat{n}_{\uparrow})/n_{\mathrm{res}}$.

Both the trace and the number of states in the subspace grow exponentially with the size of the system $L$. One must use combinatorics to find $\langle\hat{n}_{\uparrow}\rangle_{\infty}$ as a function of $L$, then take $L \to \infty$. We do this below, although in general it is a challenging problem. Before doing so, we first show that a much more simple approach can be taken by using a rate equation formalism.

Assume that out of all the lattice sites, a fraction $N_{\uparrow}$ have an $\uparrow$ single atom, $N_{\downarrow}$ have a $\downarrow$ single atom, $N_{d}$ have a doublon, and $N_{h}$ have a hole. This has the obvious constraint that,
\begin{equation}
    N_{\uparrow}+N_{\downarrow}+N_{d}+N_{h}=1.
\end{equation}
For a filling of 1 atom per site there is a doublon for every hole,
\begin{equation}
    N_{d} = N_h.
\end{equation}
Finally, for our initial state of all atoms in $\uparrow$ with energy $\frac{\Omega}{2}$, energy conservation gives,
\begin{equation}
    \frac{\Omega}{2}N_{\uparrow}-\frac{\Omega}{2}N_{\downarrow} + U N_{d} = \frac{\Omega}{2}.
\end{equation}
Combining the above three equations yields,
\begin{equation}
\label{eq_EnergyConservation}
    N_{\uparrow} + \left(1 + \frac{U}{\Omega}\right)N_{d} = 1.
\end{equation}
For a resonance $\Omega = \frac{1}{q}U$ we then have,
\begin{equation}
    N_{\uparrow} + \left(1 + q\right)N_{d} = 1.
\end{equation}

The resonant process creates one doublon from $q+1$ adjacent $\uparrow$ singlons. We can write this relation as a rate equation,
\begin{equation}
    \frac{d}{dt} \left(\begin{array}{c} N_{\uparrow} \\ N_{d}\end{array}\right) \sim \left(\begin{array}{cc} -(q+1) & (q+1) \\ 1 & -1\end{array}\right) \left(\begin{array}{c} N_{\uparrow} \\ N_{d}\end{array}\right),
\end{equation}
where the rate prefactor will be given by the respective resonant process ($J_{\perp}$ for $q=1$, $\frac{4J _{\perp}^2}{U}$ for $q=2$, etc.). At time $t=0$ we have $N_{\uparrow}=1$ and $N_{d}=0$. At equilibrium $t \to \infty$ the time derivative is equal to zero, which yields a solution of,
\begin{equation}
    N_{\uparrow}(t \to \infty) = N_{d}(t \to \infty) = \frac{1}{q+2}.
\end{equation}
Since both an $\uparrow$ singlon and a doublon contain a $\uparrow$ atom, the equilibrium value of the $\langle \hat{n}_{\uparrow}\rangle$ observable can be predicted as:
\begin{equation}\label{eqn: rate eqn soln}
    \frac{\langle\hat{n}_{\uparrow}\rangle_{\infty}}{L} = N_{\uparrow}(t \to \infty) + N_d (t \to \infty)= \frac{2}{q+2}.
\end{equation}
For resonances $\Omega = U$, $U/2$, $U/3$ this yields $2/3$, $1/2$, $2/5$ respectively as discussed in the main text. The values of other observables like $\langle \hat{n}_{\downarrow}\rangle_{\infty}$ can be predicted using the conservation relations above.

We note that this simple, coarse-grained approach relies on the assumption that the entire accessible subspace can be reached with resonantly-allowed moves. This assumption is true for our simple initial state of all-$\uparrow$ singlons. However, it may not hold for a generic initial state, especially in special regimes such as $\phi = \pi$ for which tunneling is more constrained, or for $n>2$ systems. A local rate equation approach would be needed in such a case to account for the local constraints, and quantum interference effects may become more relevant.

Having shown a simple way to predict the thermal average, we now directly compute the microcanonical ensemble average by counting the accessible many-body states of the system. A many-body state has a total number of $\uparrow$ singlons $\mathcal{N}_{\uparrow} = L N_{\uparrow}$, $\downarrow$ singlons $\mathcal{N}_{\downarrow} = L N_{\downarrow}$, doublons $\mathcal{N}_{d}= L N_{d}$ and holes $\mathcal{N}_{h}=L N_{h}$. For any given set of numbers $\{\mathcal{N}_{\uparrow}, \mathcal{N}_{\downarrow},\mathcal{N}_{d},\mathcal{N}_{h}\}$ there are $L! /(\mathcal{N}_{\uparrow}! \mathcal{N}_{\downarrow}! \mathcal{N}_{d}! \mathcal{N}_{h}!)$ ways to permute them among the lattice sites, and hence that many different many-body states. However, using the conservation relations we can reduce these numbers to just the number of doublons,
\begin{equation}
\begin{aligned}
    \mathcal{N}_{\uparrow} &= L-(q+1)\mathcal{N}_{d},\\ 
    \mathcal{N}_{h} &= \mathcal{N}_{d},\\
    \mathcal{N}_{\downarrow} &= (q-1)\mathcal{N}_{d}.
\end{aligned}
\end{equation}
To ensure that $\mathcal{N}_\uparrow, \mathcal{N}_h, \mathcal{N}_\downarrow \geq 0$, the number of doublons $\mathcal{N}_d$ can take on values of $\mathcal{N}_{d} = \{0,1,\dots \frac{L}{1+q}\}$. The total number of states in the accessible subspace is then,
\begin{equation}\label{eqn: n_res}
\begin{aligned}
    n_{\mathrm{res}} &= \sum_{\mathcal{N}_{\uparrow},\mathcal{N}_{\downarrow},\mathcal{N}_{d},\mathcal{N}_{h}} \frac{L!}{\mathcal{N}_{\uparrow}! \mathcal{N}_{\downarrow}! \mathcal{N}_{d}! \mathcal{N}_{h}!}\\
    &= \sum_{\mathcal{N}_{d}=0}^{L/(1+q)} \frac{L!}{\mathcal{N}_{d}!^2\left[(q-1)\mathcal{N}_d\right]! \left[L-(q+1)\mathcal{N}_{d}\right]!}.
\end{aligned}
\end{equation}
Furthermore, the trace of the observable can also be obtained since the expected number of $\uparrow$ atoms for each state with a given $\mathcal{N}_{\uparrow}$ is $(\mathcal{N}_{\uparrow}+\mathcal{N}_{d})/L = (L-q\mathcal{N}_{d})/L$,
\begin{equation}\label{eqn: Tr P_res n_res}
\begin{aligned}
    &\text{tr}(\hat{P}_{\mathrm{res}}\hat{n}_{\uparrow}) = \\
    &\sum_{\mathcal{N}_{d}=0}^{L/(1+q)} \frac{L!}{\mathcal{N}_{d}!^2\left[(q-1)\mathcal{N}_d\right]! \left[L-(q+1)\mathcal{N}_{d}\right]!} \frac{L-q \mathcal{N}_{d}}{L}.
\end{aligned}
\end{equation}

The sums in Eq.~\eqref{eqn: n_res} and~\eqref{eqn: Tr P_res n_res} can be explicitly evaluated, and tend to involve generalized Hypergeometric functions. For example, $n_{\rm res} = {}_2F_1\left(\frac{1-L}{2}, \frac{-L}{2}, 1; 4\right)$ for $q=1$. However, a much simpler way to approximate the ratio of the sums in the $L \to \infty$ limit is to replace the sums by the maximum values of their respective summands, which occur at $\mathcal{N}_d = \frac{L}{2+q}$ for both sums. Replacing the sums by the summand at $\mathcal{N}_d = \frac{L}{2+q}$ gives
\begin{equation}
\frac{ \braket{ \hat{n}_\uparrow }_{\infty} }{L} = \frac{\text{tr}\left(\hat{P}_{\mathrm{res}}\hat{n}_{\uparrow}\right)}{n_{\mathrm{res}}} \approx \frac{L-q\mathcal{N}_d}{L} = \frac{2}{q+2}.
\end{equation}
This value of $\frac{ \braket{ \hat{n}_\uparrow }_{\infty} }{L}$ agrees with the value we predicted from the rate equations in Eq.~\eqref{eqn: rate eqn soln}.

\end{appendix}

\end{document}